\documentstyle[12pt,epsfig]{article}
\begin{document}
\def\bfg #1{{\mbox{\boldmath $#1$}}}
\eject
\begin{center}
{\large \bf {
 STRUCTURE OF THE $~^3He$ IN BACKWARD ELASTIC $p~^3He$-SCATTERING
   }}
\vskip 1em
 Yu.N. Uzikov
\footnote{e-mail address: uzikov@nusun.jinr.dubna.su\\
Permanent address: Department of Physics, Kazakh State
 University, Tole bi 96, Almaty 480012 Kazakhstan
}

{\it Laboratory of Nuclear Problems,
 Joint Institute for Nuclear Research,\\
 Dubna, Moscow reg., 141980 Russia}
\end{center}

\section*{ Abstract}
 Backward elastic $p~^3He$-scattering at incident  proton kinetic
 energies  $T_p>1$ GeV is investigated in the framework of
 the np-pair transfer mechanism and triangular diagram of
 one-pion exchange with a subprocess $pd\to ~^3He\pi^0$
 using a realistic three-body wave function of the $~^3He$ nucleus.
 It is found that the $np-$pair transfer mechanism dominates owing to
 a rich high momentum component of the $~^3He$ wave function.
 We  show that the experimental cross section of this process is
 defined mainly by the values  of the Faddeev component of the $~^3He$
 wave function, $\varphi ^{23}({\bf q}_{23}, {\bf p}_1)$, at high
 relative momenta  $q_{23}> 0.6 GeV/c$ of the NN-pair in the $^1S_0-$
 state  and at low spectator momenta $p_1\leq 0.1$ GeV/c. The  spin-spin
 correlation parameter is calculated in the framework of the dominating
 mechanism for the case of polarized target and beam.
 Rescatterings in the initial and final states
 are taken into account. Comparison with the $pd\to dp$ process  is
 performed.
\vskip 0.5em
PACS numbers: 25.10+s, 25.40.Cm, 21.45.+v\\
Keywords: high momentum components, $~^3He$, Faddeev wave function

\eject

\section{Introduction}

 Owing to high momentum transfers $\Delta> 1$ GeV/c
 in backward elastic scattering of protons from deuteron, $~^3He$ and
 $~^4He$ nuclei at  initial proton kinetic energies about of $1$ GeV,
 the experimental data on these processes contain in principle an
 information on the structure of the lightest nuclei at short
 NN-distances  in the  region of nucleon overlap,
 $r_{NN}\sim 1/\Delta \leq 0.5$fm.
 However theoretical analysis performed most carefully in the simplest
 case of the backward elastic pd scattering
 \protect\cite{kolybassm} - \protect\cite{kaptari98},
 has not yet given quantitative
 results  on the deuteron structure at short relative distances between
 the neutron and the proton, in particularly, on relativistic
 \cite{kaptari98} and  $NN^*$- components \cite{uz979} of the deuteron
 wave function. It seems likely that this fact is connected to the
 extremely small
 binding energy of the deuteron. As a result  the high momentum component
 of the deuteron wave function is not rich enough
 to dominate  in the amplitude  of the process $pd\to dp$.
 On the contrary,  mechanisms resulting in excitation of nucleons inside
 the deuteron due to interaction with an incident beam
 give a significant contribution to the cross section of the
  $pd \to dp$ process \protect\cite{kolybassm}-\protect\cite{naksat}.
 The $~^3He$ nucleus as a more compressed system differs essentially
   from  the deuteron
 and consequently one can obtain interesting results from  investigation
 of the   $p~^3He\to ~^3Hep$ process.

\begin{figure}{}
\mbox{\epsfig{figure=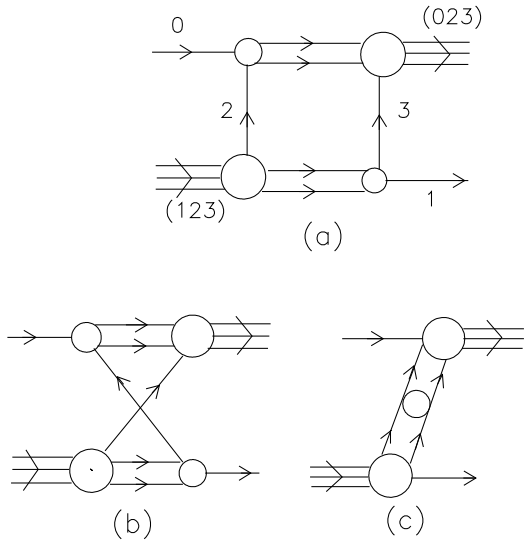,height=0.35\textheight, clip=}}
\caption{
 The np-pair transfer mechanisms of the
 backward elastic $p~^3He$-scattering denoted as $0+(123)\to 1+(023)$:
 {\it a} --  sequential transfer  (ST), {\it b} -- nonsequential
 transfer  (NST), {\it c} -- interacting pair transfer (IPT).
}
\label{fig1}
\end{figure}

  The cross section of backward elastic $p~^3He$-scattering at
 the kinetic energy  of incident proton $T_p> 1$ GeV displays three
 remarkable peculiarities \cite{laduz92,[2]}.
 (i) In the Born approximation
 only one mechanism of the process $p^3He\to ~^3Hep$ dominates, it is the
 so-called sequential  transfer (ST) of the noninteracting np-pair
 (Fig.\ref{fig1}).
 The contribution from the  mechanisms of nonsequential transfer (NST),
 interacting np-pair  transfer (IPT) and deuteron exchange  is negligible.
 In Refs.  \cite{gurvitz} -\cite{sherif}
 the heavy particle stripping  ( {\it et id.} two-nucleon transfer)
 was also investigated  and found to be important at back angles
  for $T_p\leq 0.6$ GeV. However the phenomenological $^3He$ wave functions
  restricted to  the two-body configuration, which does not permit
  ST-mechanism  were used in that analysis. The other group of papers
\protect\cite{pl,ls} based on  the microscopic optical potential constructed
 using antisymmetrized $pN$-amplitudes gives a qualitative
explanation of a rise of the cross section at backward angles at energies
$T_p\leq 0.6$ GeV without taking into account the heavy particle stripping
 mechanism. An application of that method at higher energies
 $T_p> 1$GeV is very complicated due to lack of the experimental
 information about the elastic formfactor of $~^3He$ in the relevant region
 of the $\Delta$ variable.
 (ii) The channel $\nu=1$ (in the notation of Ref.\cite{bkt})
 of the   Faddeev component $\varphi^{23}({\bf q}_{23}, {\bf p}_1)$
 of the $^3He$  wave function  plays the  most  important role in the
 $np$-pair transfer mechanism.  This  channel  corresponds
 to the orbital momentum $L=0$, spin $S=0$, isotopic spin $T=1$ of two
  nucleons with numbers 2 and 3 and the orbital momentum $l=0$ of
 the nucleon spectator denoted by the number 1.
 If this channel is excluded from the full wave function
 $\Psi=\varphi^{23}+ \varphi^{31}+\varphi^{12}$, the cross section falls
 by several orders of magnitude.
 (iii) Rescatterings in the initial and final
 states decrease the cross section
 at  $\theta_{c.m.}=180^0$  considerably in comparison with the Born
 approximation and make it agree satisfactorily with
 the available experimental data \cite{berth} for $T_p> 0.9$ GeV.

\begin{figure}{}
\mbox{\epsfig{figure=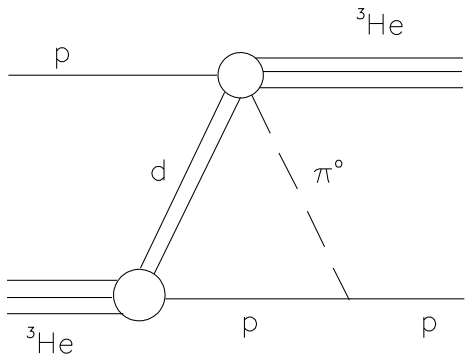,height=0.35\textheight, clip=}}
\caption{
 The triangular diagram of one-pion exchange (OPE)
  with the subprocess  $pd\to ~^3He\pi ^o$  for backward elastic
  $p~^3He$ scattering.
}
\label{fig2}
\end{figure}

 Owing to this evident connection between the structure of $~^3He$
 nucleus and the dominating mechanism one can hope to obtain an information
 about high momentum components of the $~^3He$ wave function from the cross
 section  of the $p^3He\to ~^3Hep$ process. However, in Refs.
 \cite{laduz92,[2]} it was mentioned that the D-components of $^3He$ wave
 function  are of surprisingly
 minor importance in the process under discussion at $T_p> 1$ GeV. Moreover,
 relativistic effects estimated in Ref. \cite{[2]} at $T_p\sim 1$ GeV
 by means of substituting the relativistic arguments into the $~^3He$ wave
 function instead of the nonrelativistic ones give rather small contribution
 to the cross section.
 For this reason, in Refs. \cite{laduz92,[2]} it was  concluded  that
 the sensitivity of the  $p~^3He\to ~^3Hep$ cross section to the high
 momentum components of the $~^3He$ wave function is rather weak in spite of
 high  momenta transferred at $T_p>1$ GeV.
 Moreover, as was found in \cite{berth},
 the role of the triangular diagram of one-pion exchange (OPE) with the
 subprocess  $pd\to ~^3He\pi^o$ related to the $\Delta$ - and double
 $\Delta$-excitation is in qualitative agreement
  with the absolute value of the experimental cross section at
 $T_p> 0.5$ GeV.

  In the present  work it is shown  that  the absolute value
 of the  $p~^3He\to ~^3Hep$ cross section at  $\theta_{c.m.}=180^o$
 and $T_p>1$ GeV  is determined mainly by the high momentum component
 of the Faddeev S-wave function of $^3He$, $\varphi^{23}({\bf q}_{23},
 {\bf p}_1)$, associated with the relative momentum ${\bf q}_{23}$.
 On the other hand,  rather low  values of  the  "spectator" momentum
 ${\bf p}_1$ are involved in the amplitude of this process.
 The cross section is calculated also in framework  of the OPE mechanism.
 The relative contribution of the OPE mechanism to the cross section
 is found to be small in comparison with the np-pair transfer mechanism.
 We show that this result is directly  connected
 to the high momentum component presented in the $~^3He$ wave function.
 It is shown also that
 due to rescatterings in the initial and final states the
 cross section calculated with
 the OPE mechanism is an order of magnitude lower in comparison with
 the experimental data.

   The paper is organized in the following way.
 Some elements of the formalism for the np-pair transfer mechanism  and
 the OPE amplitude
 are presented in the next section. The formulas for the $~^3He$ charge
 formfactor are also derived there for the channel
 $\nu=1$ of the three-body wave function of  $~^3He$
 and for the $d+p$ configuration. Numerical results and
 discussions are given in the section 3. The detail formulas for the np-transfer
 mechanism in the S-wave approximation are presented in the Appendix
 with the analytical gaussian parametrization of the wave function  and the
 corresponding numerical parameters are given there.

\section{Formalism}
\subsection{np-pair transfer mechanism}

  The total formalism for the np-pair transfer mechanism of the backward
 elastic $p~^3He$-scattering was developed in detail in Refs.
 \cite{laduz92,[2]}.
 We use here this formalism in the particular case of the S-wave
 component of the $~^3He$  wave function.

 In the Born approximation the amplitude of  transfer of two nucleons
   with numbers 2 and 3 in the process $0+\{123\}\to 1+\{023\}$
 (et id. $p~^3He\to ~^3Hep$) can be written as  \cite{laduz92,[2]}
$$T_B=6(2\pi )^{-3}\int d^3q_{23}L_{23}(q_{23},p_1)
\chi^+_{p'}(1)\bigl \{\varphi^{{23}^+}_f(0;23) \varphi^{31}_i(2;31)+$$
 \begin{equation}
\label{u1}
\varphi_f^{{02}^+}(3;02) \varphi_i^{31}(2;31)+
\varphi_f^{{30}^+}(2;30) \varphi_i^{31}(2;31)\bigr \}\chi_{p}(0),
\end{equation}
 where $\varphi^{ij}(k;ij)=\varphi^{ij}({\bf q}_{ij},{\bf p}_k)$
 is the Faddeev component of the wave function of the bound
 state $\{ijk\}$, $\chi_{p}(\chi_{p'})$ is the spin -- isotopic spin
 wave function of the incident (final) proton;
 $L_{23}=\varepsilon+{\bf q}^2_{23}/m_p+3{\bf p}^2_1/4m_p$,
 $m_p$ is the proton
 mass, $\varepsilon $ is the $~^3He$ binding energy.
 The subscripts {\it i} and {\it f} in Eq.( \ref{u1}) refer to the initial
 and final nucleus respectively.
 The terms $\varphi^{{23}^+}_f \varphi^{31}_i $,
 $\varphi^{{02}^+}_f \varphi^{31}_i$,
 $\varphi^{{30}^+}_f \varphi^{31}_i$ correspond  to the IPT, ST and
 NST mechanisms respectively.   In the explicit form the ST mechanism
 has the following structure of the arguments of the wave functions
$$\varphi^{{02}^+}_f \varphi^{31}_i=
 \varphi^{{02}^+}_f
\bigl ({\bf q}_{02}=-\frac{1}{2}{\bf q}_{23}-\frac{3}{4}{\bf Q}_0,
{\bf p}_3={\bf q}_{23}-\frac{1}{2}{\bf Q}_0 \bigr )\,$$
\begin{equation}
\label{u2}
 \times\varphi^{{31}}_i
\bigl ({\bf q}_{31}=-\frac{1}{2}{\bf q}_{23}+\frac{3}{4}{\bf Q}_1,
 {\bf p}_2=-{\bf q}_{23}-\frac{1}{2}{\bf Q}_1 \bigr ),
 \end{equation}
 where ${\bf Q}_0\, ({\bf Q}_1)$ is the momentum of incident (final) proton
 in the c.m.s of the final (initial) $~^3He$ nucleus. As was noted in
\cite{[2]}, at the scattering angle $\theta_{c.m.}=180^o$ two of four momenta
 in Eq.(\ref{u2}) can simultaneously become equal to zero at integration over
 ${\bf q}_{23}$. On the contrary, in the corresponding formulas for the
 IPT and NST mechanisms only one argument can be equal to zero while
 the other three have large values $\sim |{\bf Q}_1|= |{\bf Q}_0|$
 (see Appendix).
 This makes the ST term dominate  in Eq.(\ref{u1}).
 Indeed, the ST mechanism takes place only if the channels with the isotopic
 spin $T=1$ of the pair of nucleons \{ij\} are included into the component
 $\varphi^{ij}(ij;k)$ either in the initial or final state.
 It is the direct consequence of the
 fact that the ST diagram in Fig.\ref{fig1}, {\it a} either starts with
 or ends in the pp-interaction.
 The $~^3He$ wave function  from Ref. \cite{bkt} contains only one
 such channel ($\nu=1$), namely, with the $^1S_0$ state of the NN-pair.
 In the S-wave approximation for the $~^3He$ wave function the cross
 section decreases by  5-6 orders of magnitude for $T_p>1$ GeV if the channel
 $\nu=1$ is excluded \cite{luz93}. The channels with $\nu \not = 1$
 corresponding to the isotopic spin $T=0$ of the NN-pair (in particularly,
 the D-components) can enter the ST-amplitude only in
 combination with the channel $\nu=1$. For this reason the role of those
 channels is not so significant.

  Since the contribution of the interacting pair transfer mechanism is
 insignificant \cite{[2]} we will discuss here only the amplitude of
 noniteracting pair transfer
(NPT) which is the sum of ST and NST amplitudes, IPT=ST+NST.
 On the basis of the formalism  \cite{[2]} the spin structure of the NPT
 amplitude in the S-wave
 approximation for the $~^3He$ wave function
  can be written in the following form
$$T_B^{\sigma ' m'\,\sigma m}( {NPT})=-6(2\pi )^{-3}\sum _
{\nu ,\nu'=1,2}\int
d^3q_{23} L_{23}({\bf q}_{23},{\bf p}_1)$$
\begin{equation}
\label{spincom}
  \times \Phi_{\nu'}({ q}_{02}, { p}_3')\left [
 \Phi_{\nu}({ q}_{12}, { p}_3)\, a_{S'S}^{\sigma ' m'\, \sigma m}
- \Phi_{\nu}({ q}_{31}, { p}_2)\, b_{S'S}^{\sigma ' m'\, \sigma m}\right ],
\end{equation}
 where $S(S')=0$ for $\nu(\nu')=1$
 and $S(S')=1$ for $\nu(\nu')=2$; $\sigma (\sigma'$)
 and $m(m')$ are the spin z-projections  of the initial
 (final) proton and the $~^3He$ nucleus, respectively;
\begin{eqnarray}
\label{ab}
a_{S'S}^{\sigma ' m'\,\sigma m}=\left (f_{S'S0}^{\sigma ' m'\,\sigma m}
+f_{S'S1}^{\sigma ' m'\,\sigma m}\right )\left (
 A_{T'0}A_{T0}+ A_{T'1}A_{T1}\right ),\nonumber \\
b_{S'S}^{\sigma ' m'\,\sigma m}=\left (f_{S'S0}^{\sigma ' m' \,\sigma m}
+f_{S'S1}^{\sigma ' m' \,\sigma m}\right )\left (
 A_{T'0}A_{T0} - A_{T'1}A_{T1}\right ),
\end{eqnarray}
here
\begin{eqnarray}
\label{fss}
f_{S'S{\widetilde S}}^{\sigma ' m'\,\sigma m} = (2{\widetilde S}+1) 3
\sum _M ({\widetilde S}M\frac{1}{2}\sigma|\frac{1}{2} m')
({\widetilde S}M\frac{1}{2}\sigma '|\frac{1}{2} m) A_{S'{\widetilde S}}
A_{S {\widetilde S}},\\
A_{S'{S}} =\sum_{j=0,1} (-1)^j \, (2j+1)
W(\frac{1}{2}\frac{1}{2}\frac{1}{2} \frac{1}{2};{ S}j)
W(\frac{1}{2}\frac{1}{2}\frac{1}{2} \frac{1}{2};{ S}'j),
\end{eqnarray}
$A_{T'T}$ is defined similarly to $A_{S'S}$.
The Jacobi relative  momenta can be written as
\begin{eqnarray}
\label{momenta}
{\bf q}_{02}=-\frac{1}{2}{\bf q}_{23}-\frac{3}{4}{\bf Q}_0, \ \ \ \
{\bf p}'_{3}={\bf q}_{23}-\frac{1}{2}{\bf Q}_0 ,\nonumber \\
{\bf q}_{31}=-\frac{1}{2}{\bf q}_{23}+\frac{3}{4}{\bf Q}_1, \ \ \ \
{\bf p}_{2}=-{\bf q}_{23}-\frac{1}{2}{\bf Q}_1,\nonumber \\
{\bf q}_{12}=-\frac{1}{2}{\bf q}_{23}-\frac{3}{4}{\bf Q}_1, \ \ \ \
{\bf p}_{3}={\bf q}_{23}-\frac{1}{2}{\bf Q}_1.
\end{eqnarray}
In the nonrelativistic case the momenta ${\bf Q}_1$ and  ${\bf Q}_0$
are expressed by the momenta of the observed particles as
\begin{eqnarray}
\label{q1q0}
{\bf Q}_1=\frac{1}{3}{\bf p}'_h -{\bf p}, \ \ \
{\bf Q}_0=\frac{1}{3}{\bf p}_h -{\bf p}',
\end{eqnarray}
 where $ {\bf p}_h ( {\bf p}'_h)$ is the momentum
 of the initial  (final)  $~^3He$ nucleus and $ {\bf p} ( {\bf p}')$
 is the momentum of the initial (final) proton in the $p+~^3He$ c.m.s.

 Performing summation over the channels
  $\nu, \nu '=1,2$ one obtains
$$T_B^{\sigma ' m'\,\sigma m}(NPT)=-6(2\pi )^{-3}
\Bigl \{ M_0\delta_{\sigma m'}\delta_{\sigma 'm}+ \sum_{\lambda }
(1\lambda \frac{1}{2}\sigma |\frac{1}{2} m')$$
\begin{equation}
\label{spinm1m2}
 \times (1\lambda \frac{1}{2}\sigma' |\frac{1}{2} m) M_1\Bigr \},
\end{equation}
\begin{equation}
\label{m0}
M_0=\frac{1}{24}\left (5 I_{33}^{11} -4I_{32}^{11}\right )+
\frac{3}{8}I_{33}^{22}+\frac{1}{8}
\left ( I_{33}^{12} -2I_{32}^{12}\right )+
\frac{1}{8}\left ( I_{33}^{21} -2I_{32}^{21}\right ),
\end{equation}
\begin{equation}
\label{m1}
M_1=\frac{1}{8}\left (5 I_{33}^{11} +4I_{32}^{11}\right )+
\frac{1}{8}I_{33}^{22}-\frac{1}{8}
\left ( I_{33}^{12} +2I_{32}^{12}\right )-
\frac{1}{8}\left ( I_{33}^{21} +2I_{32}^{21}\right ),
\end{equation}
\begin{equation}
\label {int}
I^{\nu '\nu}_{kl}\equiv I^{\nu '\nu}_{(0i)k;(j1)l}=
\frac{1}{(4\pi )^2}n_{\nu '}n_{\nu }\int d^3q_{23}
L({\bf q}_{23}, {\bf p}_1)
\Phi_{\nu'}({q}_{0i},{ p}_k)
\Phi_{\nu}({q}_{j1},{ p}_l).
\end{equation}
\eject
 The integral $I^{\nu '\nu}_{33}$ corresponds to the NST-mechanism
 whereas $I^{\nu '\nu}_{32}$ describes the ST mechanism. The amplitude
 $T_B$ in Eq. (\ref{u1}) is related to the invariant amplitude $A$ by the
 following
 formula
\begin{equation}
\label{sect}
A_B=4m_pm_h T_B,
\end{equation}
 the connection of this amplitude with the cross section is given in
 the next section by Eq.(\ref{a2sec}). The total wave function of
the $~^3He$ nucleus
$\Psi=\varphi^{12}+\varphi^{23}+\varphi^{31}$
 in Eq. (\ref{u1}) is  normalized as
\begin{equation}
\label{norma3he}
\int\left | \Psi({\bf q},{\bf p})\right |^2\frac {d^3 q\, d^3p}
{(2\pi )^6}=1.
\end{equation}

\subsection {Spin-spin correlation parameter}

   The spin-spin correlation parameter $\Sigma $ is calculated here
as an additional test of the np-transfer mechanism of the process
${\vec p} ~^3{\vec He}\to ~^3He p$ with polarized beam
 and target. This parameter is defined as
\begin{equation}
\label{u5}
\Sigma=\frac{d\sigma(\uparrow \uparrow)-d\sigma(\uparrow \downarrow)}
 {d\sigma(\uparrow \uparrow)+d\sigma(\uparrow \downarrow)},
\end{equation}
 where $d\sigma(\uparrow \uparrow)/d\Omega$ and
$d\sigma(\uparrow \downarrow)/d\Omega$ are
the cross sections for  parallel and antiparallel spins of colliding
 particles, respectively.
 From Eq. (\ref{spinm1m2}) follows that the sum of ST and NST amplitudes
 in the S-wave approximation is not zero only  for
  $m'+\sigma '=m+\sigma$. Consequently from 16 amplitudes given by
 Eq.(\ref{spincom}) the nonvanishing  ones are the following
\begin{eqnarray}
T_1\equiv T^{++++} = T^{----}= M_0+\frac{1}{3}M_1,\nonumber \\
T_2\equiv T^{+-+-}=T^{-+-+}=\frac{2}{3}M_1,\nonumber \\
T_3\equiv T^{-++-}=T^{+--+}=M_0-\frac{1}{3}M_1.
\label{t178}
\end{eqnarray}
From Eqs. (\ref{u5}) and (\ref{t178}) one finds
\begin{equation}
\Sigma= \frac{|T_1|^2 -|T_2|^2-|T_3|^2}{|T_1|^2 +|T_2|^2+|T_3|^2}=
\frac{2}{3}\frac{Re(M_1\, M_0^*)-\frac{1}{3}|M_1|^2}
{|M_0|^2+\frac{1}{3}|M_1|^2}.
\end{equation}

\subsection{ The OPE mechanism}

   An obvious modification of the formalism of the triangular OPE diagram from
 Ref. \cite{naksat} is used here for the OPE amplitude. According to common
 rules of the diagram technique the amplitude corresponding to the triangular
 diagram in Fig. \ref{fig2} takes the following form
\begin{eqnarray}
\label{opo1}
A_{OPE}^{\sigma ' m'\,\sigma  m}(ph \to ph )=
\int \frac{d^3p_ddT_d}{(2\pi)^4}\sum_{\lambda_d \sigma_p}
A_{\sigma \lambda_d}^{m'}(pd\to ~^3He\pi^0)\nonumber \\
\times\frac{G_m^{\lambda_d \sigma_p}(~^3He\to d+p)<\pi^0 p|p'>}
 {(k^2-\mu^2+i\epsilon)(2m_pT_p-{\bf p}^2_p+i\epsilon)
 (2m_dT_d-{\bf p}^2_d+i\epsilon)},
 \end{eqnarray}
 where  ${\bf p}_i$ is the momentum, $T_i$ is the kinetic energy and
$m_i$ is the mass
 of the i-th intermediate particle
 (proton or deuteron); $k$ and $\mu $ are the 4-momentum and mass
 of the virtual  $\pi $- meson, respectively.
  In Eq.(\ref{opo1}) the summation refers to the
 spin states of the intermediate deuteron $(\lambda_d)$ and proton
 $(\sigma_p)$. The invariant amplitude
 $A$ of the process $ab\to cd$ in Eq. (\ref{opo1}) is related to the
 corresponding  differential cross section in the  c.m.s. by the
 following formula
\begin{equation}
\label{a2sec}
\frac{d\sigma}{d\Omega }= \frac{1}{64\pi ^2s_{ab}}\frac{q_{cd}}{q_{ab}}
{\overline{|A|^2}},
\end{equation}
where $s_{ab}$ is the square of the invariant mass of the system $a+b$,
$q_{ij}$ is the relative momentum in the system
 $i+j$. The amplitude of the virtual decay
 $ ~^3He\to d+p$ has the form
\begin{equation}
\label{opo2}
G_m^{\lambda_d \sigma_p}(~^3He\to d+p)=\sqrt{S_{pd}^{h }}\,4m_p\sqrt{m_h}(\varepsilon_h+
\frac {2}{3}\frac{{\bf Q}^2}{m_p})\psi_m^{\lambda _d \sigma_p}({\bf Q});
\end{equation}
 here $ m_h $ is the mass of the $~^3He$ nucleus,
 $S_{pd}^{h}\simeq 1.5$ is the spectroscopic factor of
  $~^3He$ in the channel $d+p$ \cite{sciavilla};
 $\psi_m^{\lambda _d \sigma_p}({\bf Q})=<~^3He|d,p>$ is the overlap
integral between the $~^3He$ wave function
$\Psi_m$ and the production of
the wave functions of deuteron $\psi _{\lambda_d}$ and proton
$\varphi_{\sigma_p}$. This wave function is  normalized by the
 following condition
\begin{equation}
\label{normaq}
\frac{1}{2J_h+1}\sum_{m, \, \lambda_d, \sigma_p} \int
|\psi_m^{\lambda _d \sigma_p}({\bf Q })|^2\frac{d^3Q}{(2\pi)^3}=1.
\end{equation}
The Fourier-transformation of this function is given by
\begin{equation}
\label{furje}
\psi_m^{\lambda _d \sigma_p}({\bf Q })=\int \exp{(-i{\bf Qr})}
\psi_m^{\lambda _d \sigma_p}({\bf r })d^3r,
\end{equation}
where the wave function in the coordinate space has the following form
$$\psi_m^{\lambda _d \sigma_p}({\bf r })=<\varphi_{\sigma_p}\,
\psi _{\lambda_d}({\bfg \rho})|\Psi_m ({\bfg \rho}, {\bf r})>=$$
\begin{equation}
\label{overlap}
=\sum_{L,M, S,M_S}(L\,M\,S\,M_S|\frac{1}{ 2}\,m)
(1\lambda_d \frac {1}{2}\sigma_p|SM_S) \, U_L(r)
\, Y_{LM}( {\hat {\bf r}});
\end{equation}
 here   $S=3/2$ for $L=2$   and  $S=1/2$  for $L=0$.
The spherical functions $Y_{LM}$ and Clebsh-Gordan coefficients are used
in Eq.(\ref{overlap}) in the standard notations.
The $S-$  and $D-$components of the wave function
 $U_L(r)$ in Eq. (\ref{overlap}) were obtained in Ref.\cite{santos}
by numerical solution of the Faddeev equations with the NN-interaction
in the form of Reid soft core (RSC). The results \cite{santos} are used
 here with the following normalization condition
\begin{equation}
\label{normawfhe3}
\int_0^\infty\left [U_0^2(r)+ U_2^2(r) \right ]r^2dr=1.
\end{equation}
Actually the three-body calculations for the normalization integral
(\ref{normawfhe3}) give the value 0.43 \cite{gibbson}. Consequently
 the contribution of the OPE diagram in Fig.
  \ref{fig2} is overestimated by the condition given in Eq.(\ref{normawfhe3}).
 However there are no experimental data at present about the differential
 cross section  of the reaction   $pd^*\to ~^3He\pi^0$, where $d^*$ is the
 singlet deuteron. Under assumption
 that the cross sections of the reactions  $pd\to ~^3He\pi^0$ and
 $pd^*\to ~^3He\pi^0$ are equal to each other, the normalization condition
 in Eq. (\ref{normawfhe3}) effectively takes into account the contribution
 of the $p+d^*$ configuration  with  $pn$-pair in the singlet state.

 The vertex function  $\pi NN$ has the following form \cite{imuz88}
\begin{equation}
\label{pinn}
<\pi^0p|p'>={2m_p}\frac{f_{\pi NN}}{\mu} \varphi_{\sigma '}^+({\bfg \sigma} {\bf Q})
\varphi_{\sigma_p}F_{\pi NN}(k^2),
\end{equation}
here $\varphi _{\sigma_p}$ and $\varphi _{\sigma'}$ are
 the Pauli spinors for nucleons, $f_{\pi NN}=1$;
\begin{equation}
\label{momq}
{\bf Q}=\sqrt{\frac{E_p+m_p}{E_{p'}+m_p}}{\bf p}_{p'}-
\sqrt{\frac{E_{p'}+m_p}{E_{p}+m_p}}
{\bf p}_p,
\end{equation}
 $E_p,\, (E_{p'})$ is the total energy of the proton $p(p')$;
\begin{equation}
\label{monopol}
F_{\pi NN}(k^2)=\frac{\Lambda_\pi^2-\mu^2}{\Lambda_\pi^2-k^2}.
\end{equation}
 For the cutoff parameter $\Lambda_\pi$ in the  monopole $\pi NN$ formfactor
 defined by Eq.(\ref{monopol})
 the value  $\Lambda_\pi=0.65$ GeV/c   is used  here \cite{lee,imuz88}.

 As is shown numerically in the next section, the contribution
 of the D-component  to the OPE cross section is negligible.
  In the S-wave approximation for the  $^3He\to d+p$ channel
 the cross section of the   $p^3He\to ^3Hep$ process can be expressed
 through the cross section of the reaction
 $pd\to ~^3He\pi^0$ in the following way
\begin{eqnarray}
\label{secswave}
\frac{d\sigma}{d\Omega_{c.m.}}=\frac{1}{64\pi ^2}\frac{1}{s_{ph}}
\overline{|A_{OPE}^{\sigma ' m'\,\sigma  m}|}=\nonumber \\
=\frac{m_p\,m_h}{2\pi}\frac{E_{p'}+m_p}{E^2_{p'}}\left ( \frac{f_{\pi NN}}
{\mu}\right )^2 \,S_{pd}^h F_{\pi NN}^2(k^2)\frac {s_{pd}}{s_{ph}}\frac{q_{pd}}
{q_{\pi h}}\frac{d\sigma}{d\Omega_{c.m.}}(pd\to ~^3He\pi^o)\nonumber \\
\times \left | i\kappa {\cal F}_0({\widetilde p})+{\cal W}_{10}
({\widetilde p},{\widetilde \delta})\right |^2,
\end{eqnarray}
where
\begin{eqnarray}
\label{formfact}
{\cal F}_l({\widetilde p})=\int _0^\infty U_l(r)\, j_l({\widetilde p}r)\,
r dr,\nonumber \\
{\cal W}_{lL}(({\widetilde p},{\widetilde \delta})=
\int_0^\infty j_l({\widetilde p}r) U_L(r)(i{\widetilde {\delta }}+1)
\exp{(-i{\widetilde {\delta }}r)}dr.
\end{eqnarray}
\begin{eqnarray}
\label{kinopo}
\kappa=2m\left (\frac{1}{E_{p'}+m}+\frac{1}{E_{p'}}\right ) |{\bf p}_{p'}|,\ \ \
{\widetilde {\bf p}}=\frac{2m}{E_{p'}}{\bf p}_{p'},\\
{\widetilde \delta}^2={\widetilde {\bf p}}^2 +
(2mT_{p'}+\mu^2-2\varepsilon_h m_p)\frac{2m_p}{E_{p'}},\\
\label{kinopolas}
k^2-\mu^2=\frac{E_{p'}}{2m_p}\left ({\widetilde {\bf p}}^2-{\widetilde \delta }^2\right );
\end{eqnarray}
in Eqs.(\ref{secswave}-\ref{kinopolas}) $E_{p'}= \sqrt{m_p^2+{\bf p}^2_{p'}}$
is the total energy,
 ${\bf p}_{p '} $ is the momentum and
 $T_{p '}$ is the kinetic energy of the secondary proton in the laboratory
 system.

  Rescatterings in the initial and final states for the OPE mechanism
 are taken into account here in the line of work \cite{[2]}  on the
 basis of Glauber-Sitenko theory. According to Ref. \cite{[2]}, the
 amplitude $A_{fi}^{dist}$ for the exchange mechanism with rescatterings
 in the  $p~^3He\to ~^3He p$ process  can be related to corresponding
 amplitude  in the Born  approximation  $A_{B}$ by the following expression
$$A_{fi}^{dist}=A_B({\bf p}'_h,{\bf p}';{\bf p}_h,{\bf p})+
{i\over4\pi
p}\int d^2q\ F_{ph}({\bf q})A_B({\bf p}'_h,{\bf p}';{\bf p}_h+{\bf q},
{\bf p}-{\bf q})$$ $$+{i\over4\pi p'}\int d^2q'\ f_{pp}({\bf q}')A_B({\bf p}_
h'-{\bf q}',{\bf p}'+{\bf q}';{\bf p}_h,{\bf p})$$
\begin{equation}
\label{rescatt}
-{1\over(4\pi)^2 p' p}\int\int d^2qd^2q'\ F_{ph}({\bf q})f_{pp}({\bf q}')
A_B({\bf p}_h-{\bf q}',
{\bf p}'+{\bf q}';{\bf p}_h+{\bf q},{\bf p}-{\bf q}).
\end{equation}
 In Eq. (\ref{rescatt}) the amplitudes $f_{pp}$ and $F_{ph}$ describe
 the elastic  $pp-$  and $p~^3He-$ forward scattering, respectively.
 The last three terms in Eq. (\ref{rescatt}) take into account rescatterings
 in the intial, final state and simultaneously in the initial and final
 states, respectively. In the spinless
 approximation the amplitude of $pN$-scattering  is  parametrized in the
 standard form \cite{sitenko}
\begin{eqnarray}
\label{fpn}
f_{pN}(q) \equiv k_{pN}A_{pN}\exp{(-B_{pN}q^2)}=
{k_{pN}\sigma_N\over4\pi}(i+\alpha_N) \exp(-{1\over2}\beta_Nq^2),\ \ \
\end{eqnarray}
 where $q$ is the momentum transferred in  $pN$-scattering, $k_{pN}$ is
 the wave vector of the nucleon in  the  $p+N$
 c.\ m.\ s., $\sigma_N$ is the total cross section of the $pN$-scattering,
 $\alpha_N $  and $\beta_N$ are the phenomenological parameters fitted to
 the experimental data on  $pN$-scattering \cite{pdg20000}.
 Using the gaussian  form for the $~^3He$ density and Eq. (\ref{fpn})
 one gets an analytical form for the amplitude  $F_{ph}$
\begin{equation}
\label{fptau}
 F_{ph}(q)=k_{ph }\sum_{k=1}^3 \,A_k^{ph }
\exp{(-B_k^{ph }q^2)};
\end{equation}
here $k_{ph}$ is the the wave vector of the nucleon in  the  $p+~^3He$
 c.\ m.\ s.;
parameters  $A_k^{ph },\, B_k^{ph }$ are expressed analytically
 through parameters of  $pN$-scattering amplitude (\ref{fpn}) and the
 oscillator radius of the gaussian form for the $~^3He$ nucleus density
 \cite{czyz}. Three terms  in Eq. (\ref{fptau}) correspond to single, double
 and triple scattering of the incident proton from nucleons of the
 $^3He$ nucleus. Since the OPE  amplitude in the Born approximation
 given by Eq.(\ref{opo1}) is a smooth
 function of the kinematic variables
 $ {\bf p}_h,\, {\bf p}'_h,
{\bf p},\, {\bf p}'$, one can take this amplitude outside of the sign
of integrals over $d^2q$ and $d^2q'$ in Eq. (\ref{rescatt})
 \footnote{ When calculating the contribution of the rescatterings to the
  np-transfer amplitude the integrations over $d^2q$ and $d^2q'$ in
 Eq.(\ref{rescatt}) are performed exactly in analytical
 form. In this case the factorization like in Eq.
  (\ref{opodist}) does not occure.}.
 In this approximation the OPE amplitude with rescatterings
 has the following form
\begin{equation}
\label{opodist}
 A_{OPE}^{dist}= D\, A_{OPE}^{\sigma ' m'\,\sigma  m},
\end{equation}
 where the amplitude  $A_{OPE}^{\sigma ' m'\,\sigma  m}$ was defined
 by Eq.(\ref{opo1}) and the distortion factor has the form
\begin{equation}
\label{distort}
 D=1+\frac{i A_{pp}}{4\beta _{pp}}+ \sum _{k}\left [ \frac{i A_k^{ph }}
{4 B_k^{ph }}- \frac{ A_{pp}A_k^{ph }}{16\beta _{pp}B_k^{ph }}
 \right ];
\end{equation}
 parameters  $A_{pp},\,\beta _{pp},\, A_k^{ph },\,B_k^{ph }$
 are defined in Eqs. (\ref{fpn},\ref{fptau}).

\subsection { The charge formfactor of  $~^3He$}

 In the nonrelativistic impulse approximation the charge formfactor
 of $~^3He$  is defined as
\begin{equation}
\label{chff}
ZF_{ch}(\Delta)=\int\int \exp{(i{\bf \Delta x})}\Psi^+(1,2,3)
{\hat \rho}_{ch}({\bf x}, {\bf r}_i)\Psi(1,2,3)d{\bf x}\prod_{i=1}^3
d{\bf r}_i,
\end{equation}
where $Z=2$ and the charge density operator has the following form
\begin{equation}
\label{plotn}
{\hat \rho}_{ch}({\bf x}, {\bf r}_i)=
\sum_{i=1}^3\Big \{ \frac{1}{2}[1+\tau_z(i)]f_{ch}^p({\bf x}- {\bf r}_i)
+\frac{1}{2}[1-\tau_z(i)]f_{ch}^n({\bf x}- {\bf r}_i)\Big \};
\end{equation}
here $ {\tau }_z$ is the Pauli matrix for the z-projection of the nucleon
 isotopic spin, ${\bf r}_i$ is the coordinate of the i-th nucleon,
$f_{ch}^N({\bf y})$ is the distribution of the nucleon charge density,
 which is related to the charge formfactor of the
 nucleon in the following way
\begin{equation}
\label{ffnuklona}
F_{ch}^N({\bf q})=\int\exp{(i{\bf y q})}f_{ch}^N({\bf y})d{\bf y};
\end{equation}
$\Psi(1,2,3)=\varphi^{23}+\varphi^{12}+\varphi^{31}$ is the antisymmetrized
 wave function of $~^3He$ nucleus,
${\bf \Delta}={\bf p}_h '-{\bf p}_h $ is the transferred  momentum.
As was mentioned above,
in the backward elastic $p~^3He$-scattering the main contribution gives
the channel $\nu=1$ of the Faddeev  wave function
 $\varphi^{23}$. The following expression can be obtained for the charge
 formfactor of the  $~^3He$ nucleus taking into account only one channel
 $\nu=1$ in the  $~^3He$ wave function
\begin{equation}
\label{chargeff}
F_{ch}(\Delta)=3(F^p+\frac{1}{2}F^n)J^{23;23}(\Delta)
+(\frac{1}{4}F^p+\frac{1}{2}F^n)J^{23;31}(\Delta);
\end{equation}
here the integrals  $J^{23;31}$ and $J^{23;23}$ are defined by the
 following formulas
\begin{eqnarray}
\label{interls}
J^{23;23}(\Delta)=\frac{1}{(4\pi)^2}\int \int d{{\bf q}}d {{\bf p}}
\Phi_{\nu'}(|{\bf q}|, |{\bf p}-\frac{2}{3}{\bf \Delta}|)
\Phi_{\nu}(|{\bf q}|, |{\bf p}|);\nonumber \\
J^{23;31}(\Delta)=\frac{1}{(4\pi)^2}\int \int d{\bf q}d {\bf p}
\Phi_{\nu'}(|{\bf q}|, |{\bf p}-\frac{2}{3}{\bf \Delta}|) \nonumber \\
\times \Phi_{\nu}(|{-\frac{1}{2}\bf q}+\frac{3}{4}{\bf p}|,
|-{\bf q}-\frac{1}{2}{\bf p}|).
\end{eqnarray}
 In the framework of the  $d+p$-configuration  for the
 $^3He$ nucleus given by Eqs. (\ref{overlap}) and (\ref{normawfhe3})
 we obtain the following expression
 for the $~^3He$ charge formfactor:
$$F_{ch}(\Delta)=\frac{1}{2}\, F_{ch}^p(\Delta)\left
\{ F_{000}\left (
\frac{2}{3}\Delta \right )+
F_{022}\left (\frac{2}{3}\Delta\right )\right \}+$$
$$+\frac {1}{2} [F_{ch}^p(\Delta)+F_{ch}^n(\Delta)]\Biggl
\{ S_s^d \left( \frac{1}{2}\Delta
\right )\Biggl [ F_{000}\left (\frac{1}{3} \Delta\right )+ $$
\begin{eqnarray}
\label{ffhe3}
+F_{022}\left (\frac{1}{3}\Delta \right ) \Biggr ]+ \frac{1}{ \sqrt{8\pi} }
S_Q^d\left( \frac{1}{2}\Delta \right )
\left [\sqrt {8}F_{220}\left (\frac{1}{3}\Delta\right )-
F_{222}\left (\frac{1}{3}\Delta\right )\right ] \Biggr \},
\end{eqnarray}
where
\begin{equation}
\label{flll}
F_{lLL'}(\Delta)=\int_0^\infty j_l(\Delta\rho )U_L(\rho )\,U_{L'}(\rho )\,
\rho ^2\, d\rho.
\end{equation}
 Here  $S_s^d (\Delta),\, S_Q^d(\Delta )$ are the scalar and
 quadrupole formfactors of the deuteron (see, for example,
 Ref. \cite{burov}). The parametrization from Ref. \cite{bilenka}
 is used here for the nucleon formfactor  $F_{ch}^N(\Delta)$.

\section{Numerical results and discussions}

 Numerical calculations for the  $np$-pair transfer mechanism are performed
 here using the $~^3He$ wave function obtained in Ref.\cite{bkt}  from
 the solution of Faddeev equations in momentum space with the
 RSC potential of the NN-interaction in the  $^1S_0$ and  $^3S_1-^3D_1$-states.
 The  separable analytical parametrization from Ref.\cite{hgs}
 is used  here for the spatial part $\Phi_\nu$ of the Faddeev component of
 the $~^3He$ wave function in Eqs.(\ref{spincom}),(\ref{int}),
(\ref{interls}).
 In the notations of Ref. \cite{hgs} it has a
 form
  \begin{equation}
 \label{u3}
 \Phi_\nu({ q}_{23},{ p}_1)
= n_\nu \varphi_\nu(q_{23})\chi_\nu(p_1),
 \end{equation}
 where $n_\nu$ is the numerical constant \cite{hgs}. The
  functions $\varphi(q) $ and $\chi(p) $ in Eq.(\ref{u3}) are
 normalized according to the following conditions
\begin{eqnarray}
\label{norma}
\int_0^\infty\varphi^2(q)q^2dq=1, \int_0^\infty\chi^2(p)p^2dp=1.
\end{eqnarray}
\begin{figure}{}
\mbox{\epsfig{figure=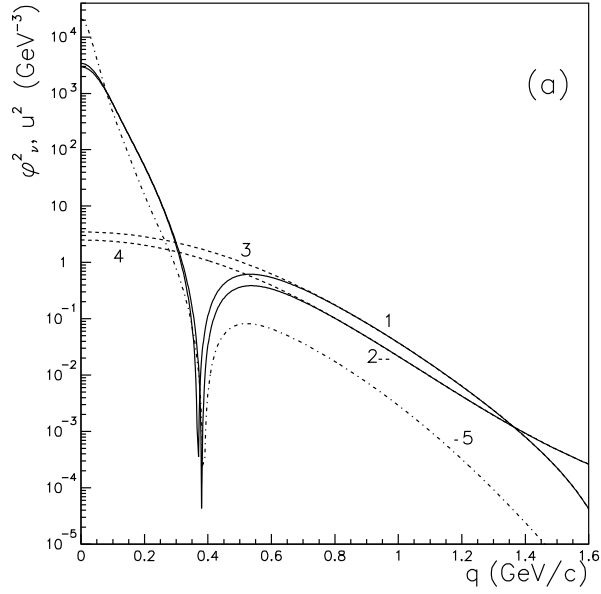,height=0.4\textheight, clip=}}
\caption{
 The square of functions
 $\varphi _\nu (q)$,  $\chi_\nu(q)$ from Ref.\protect\cite{hgs},
  the S-component of the deuteron wave function  $u(q)$
from Ref. \protect\cite{alberi} and the functions
$\widetilde  {\varphi}_\nu(q)$
 and ${\widetilde{\chi}}_\nu(q)$
 defined in the text.
 1 -- $\varphi^2 _1 (q)$,
 2 -- $\varphi^2 _2 (q)$,
 3 -- ${\widetilde{\varphi}}_1^2(q)$,
 4 -- ${\widetilde{\varphi}}_2^2(q)$,
 5 -- $u^2(q)$;
}
\label{fig3}
\end{figure}
\begin{figure}{}
\mbox{\epsfig{figure=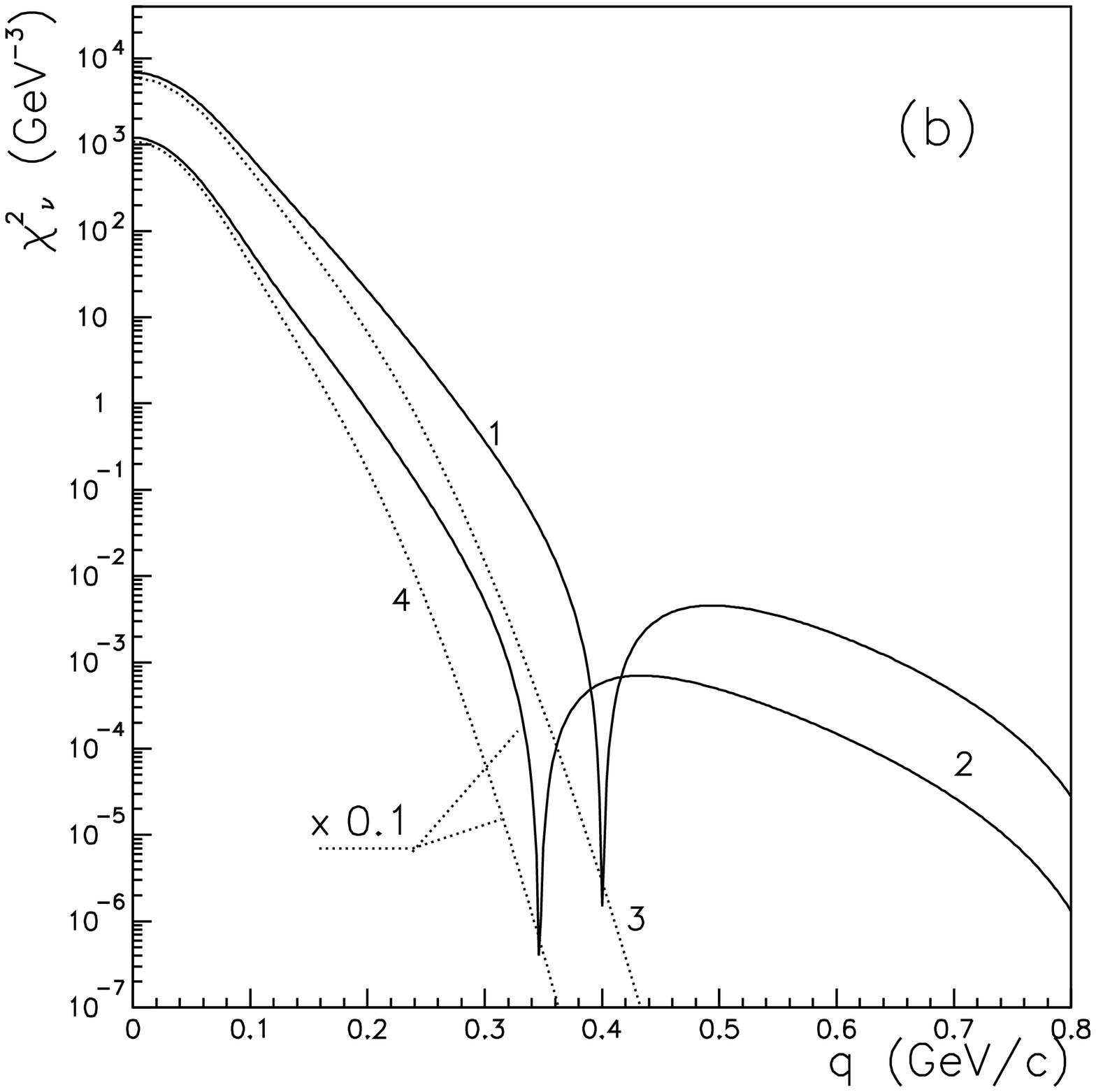,height=0.4\textheight, clip=}}
\centerline{Fig.3, b}
{\small The same as in Fig.\protect\ref{fig3},{\it a} but
 1 -- $\chi^2_1(q)$,
 2 -- $\chi^2_2(q)$,
 3 --  ${\widetilde{\chi}}_1^2(q)$,
 4 --  ${\widetilde{\chi}}_2^2(q)$; the curves 2 and 4 in
 the part {\it a} are multiplied by factor $10^{-1}$.}
\end{figure}

\begin{figure}{}
\mbox{\epsfig{figure=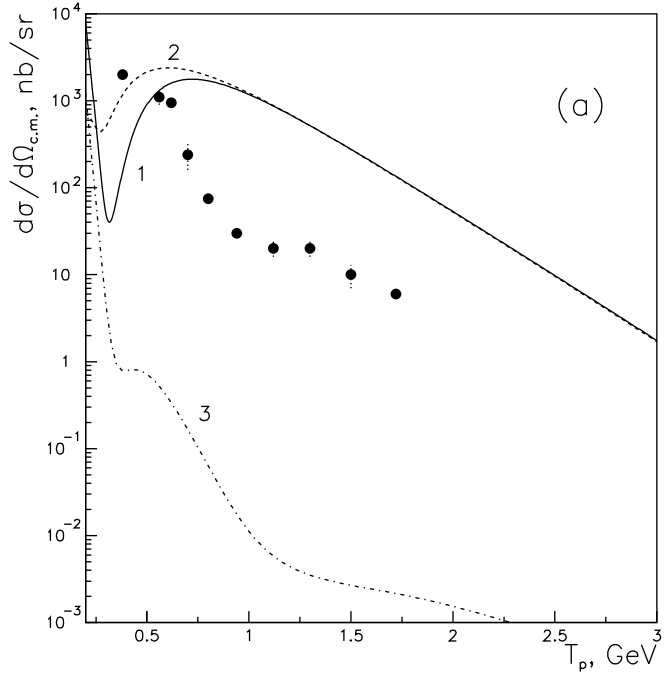,height=0.45\textheight, clip=}}
\caption{
 The differential cross section of  elastic $p^3He$ scattering
 at $\theta_{c.m.}=180^o$ as a function of the incident proton kinetic
 energy $T_p$. Curves 1-3 show the results of calculation in the
Born approximation for the amplitude in Eq. (\protect\ref{u1}):
 1 --   with the $~^3He$ wave function from \protect\cite{hgs},
 2 -- with  ${\widetilde{f}}_\nu(p_1)$ instead of ${f}_\nu(p_1)$,
 3 -- with ${\widetilde {f}}_\nu(q_{23})$  instead ${f}_\nu(q_{23})$;
{\it a} -- for $f_\nu =\varphi_1$,
{\it b} -- for $f_\nu =\chi_1$,
{\it c} -- for $f_\nu =\varphi_2$,
{\it d} -- for $f_\nu =\chi_2$.
 The experimental points are taken from Ref.\protect\cite{berth}
}
\label{fig4}
\end{figure}

 The square of the functions $\varphi_\nu(q), \chi_\nu(q)$ and
 the S-component of the deuteron wave
 function, $u(q)$, for the RSC potential \cite{alberi} are
 shown  in Fig.\ref{fig3}.  The results of  calculation of the differential
 cross section are  shown in Figs.\ref{fig4}-\ref{fig8}
 in comparison with the experimental  data  \cite{berth}.

 The  numerical results demonstrate the following important features of
 the process in question.  First,  the ST-mechanism involves the high
 momentum components  of the S-wave functions $\varphi_\nu(q_{23})$.
 The $^3He$ wave function in the
 channels $\nu=1$ and $\nu=2$ is probed at  high momenta
 ${\bf q}_{23}> 0.6 GeV$
 when the cross section is measured at $T_p> 1$ GeV. To show it,
 in Fig.\ref{fig3} ({\it a}) we present the  part  of the function
 $\varphi_\nu(q_{23})$ ($\nu=1$ and 2),
 denoted as ${\widetilde \varphi}_\nu$, which
 coincides with  $\varphi_\nu(q_{23})$ for $q_{23}>0.6 GeV/c$ and
 differs considerably    from it for smaller momenta
 $q_{23}<0.5 GeV/c$. In Fig.\ref{fig3} ({\it b}) we also show the  parts
 of the functions $\chi_\nu (p_1)$, denoted as ${\widetilde \chi}_\nu$,
 which are very close to the corresponding total functions
 $\chi_\nu (p_1)$ at small spectator
 momenta $p_1\sim 0\div 0.1 $GeV/c and are negligible for $p_1>0.2 $GeV/c.
 The cross sections calculated with these functions  ${\widetilde f}_\nu$
 instead of the  full functions  $f_\nu$ (here $f=\varphi$ or $f=\chi$)
 are shown  in Fig.\ref{fig4} by  curves 2.
 One can see that these curves are  very close
 to  the total result obtained with the full functions  $\varphi_\nu(q_{23})$
 and  $\chi_\nu (p_1)$. In contrast, as one can see from Fig.\ref{fig4}
 (curves 3), the cross section
 calculated with the  complementary  parts
 $\varphi_\nu-{\widetilde \varphi}_\nu$
 and $\chi_\nu-{\widetilde \chi}_\nu$ is 5-6 orders of magnitude smaller
 for $\nu=1$  and 4-6 times smaller for the channel $\nu=2$.
 Obviously the  channel $\nu=1$ of the $~^3He$ wave function plays the most
 important role in the   $p~^3He\to ~^3Hep$ process.

\begin{figure}{}
\mbox{\epsfig{figure=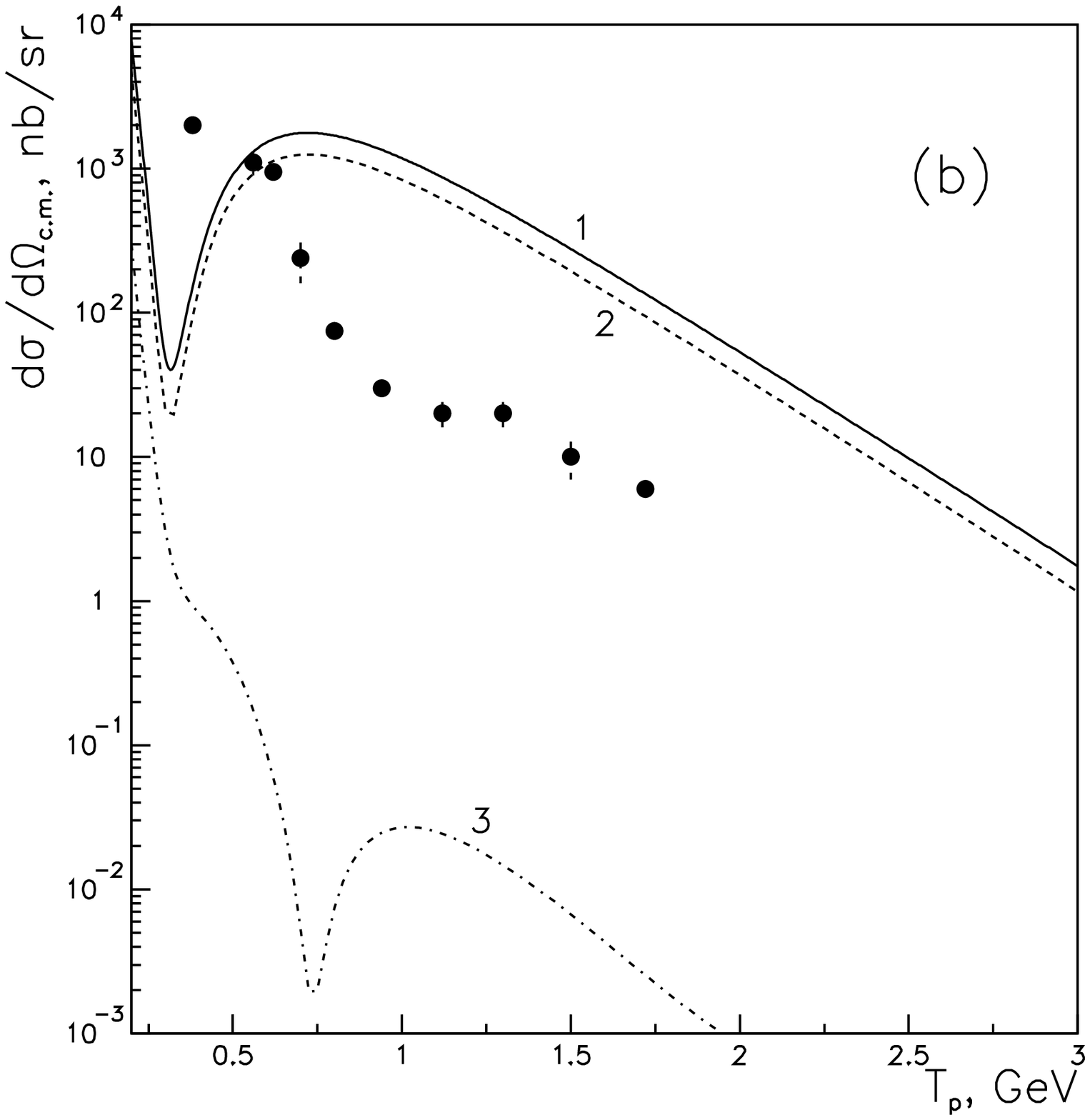,height=0.4\textheight, clip=}}
\centerline{Fig.4, b}
\end{figure}
\eject
\begin{figure}{}
\mbox{\epsfig{figure=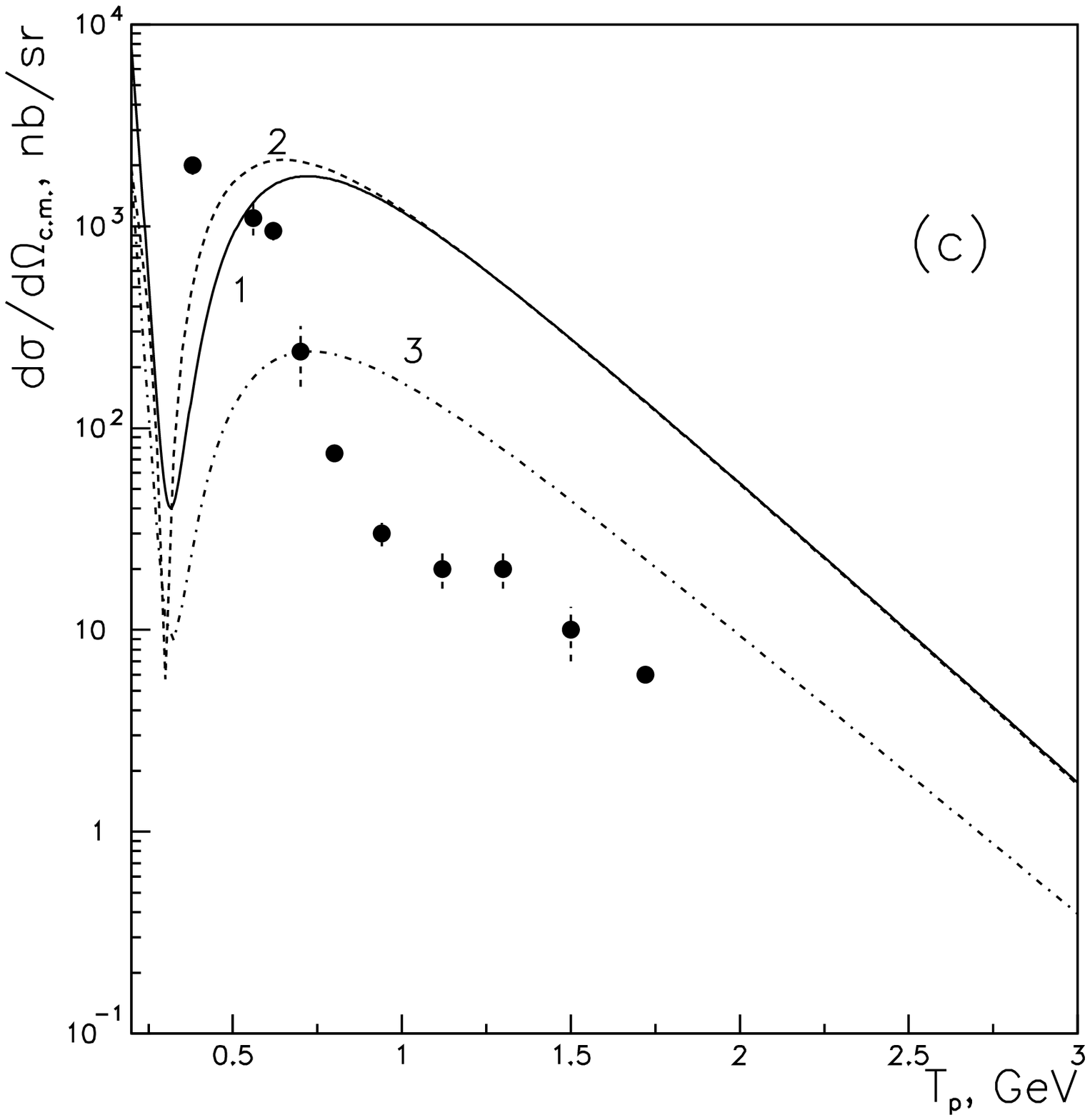,height=0.4\textheight, clip=}}
\centerline{Fig.4, c}
\end{figure}
\eject
\begin{figure}{}
\mbox{\epsfig{figure=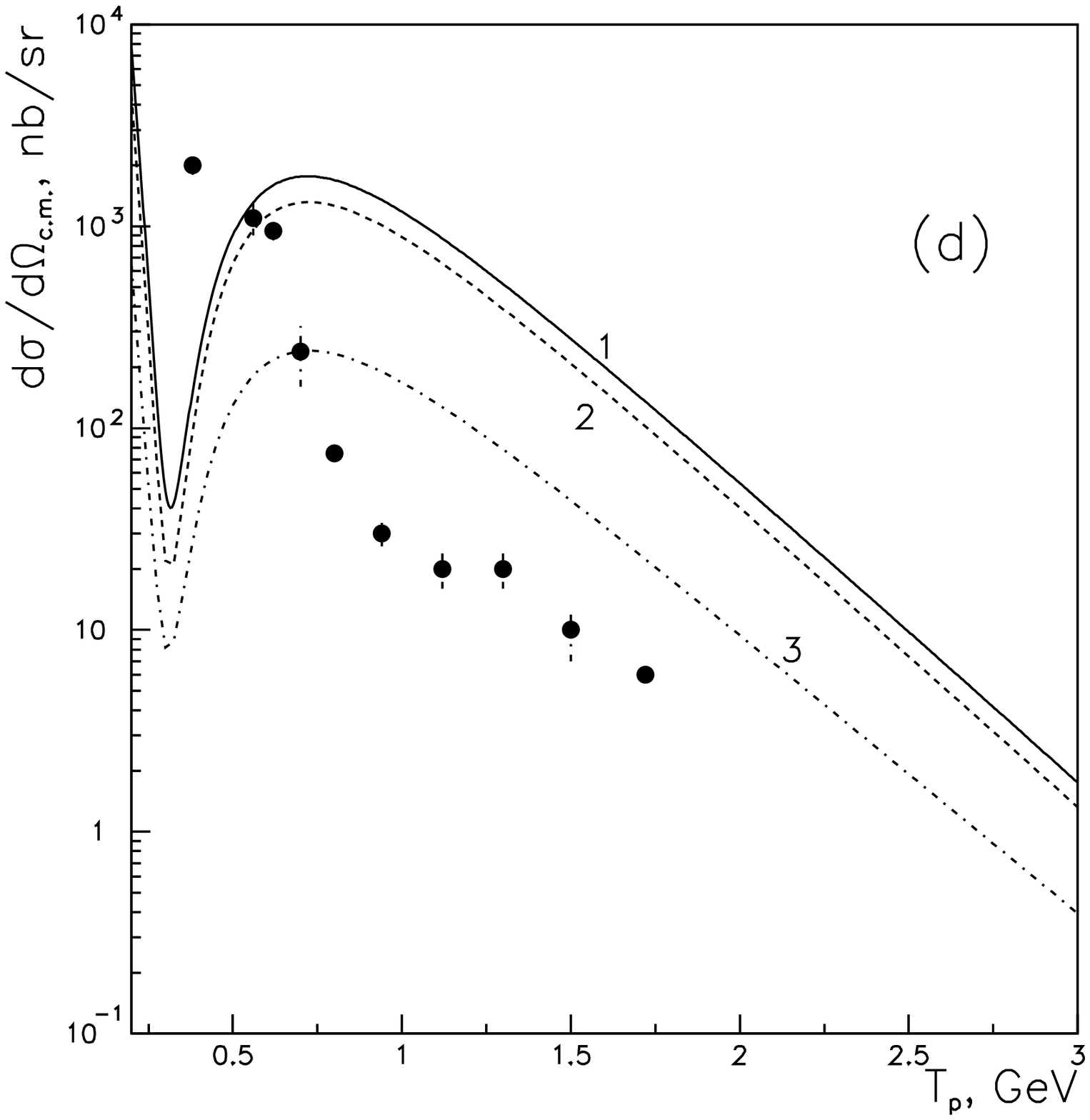,height=0.4\textheight, clip=}}
\centerline{Fig.4, d}
\end{figure}

  Second, the above  result  demonstrates also that the ST mechanism
  uses rather low  "spectator"-momenta $p_1\sim 0\div 0.1GeV/c$ in the
 function $\chi_\nu(p_1)$. This feature  makes the ST  mechanism
 dominate.  The qualitative explanation of these results is following.
 One can find from Eq.(\ref{u2}),  that for ${\bf Q}_1=-{\bf Q}_0$
({\it et id.}
 $\theta_{c.m.}=180^o$) the equations ${\bf q}_{31}={\bf q}_{02}$ and
 ${\bf p}_{2}=-{\bf p}_{3}$ are satisfied. Consequently, the main
 contribution into the  integral over $d{\bf q}_{23}$ in Eq. (\ref{u1})
 gives the region $|{\bf p}_{2}|=|{\bf p}_{3}|\sim 0$, in which
 $|{\bf q}_{31}|=|{\bf q}_{02}| \sim Q_1$. On the contrary,
 the region of $|{\bf q}_{31}|=|{\bf q}_{02}| \sim 0$ corresponds to
 $|{\bf p}_{2}|=|{\bf p}_{3}|\sim 2Q_1$ and plays insignificant role since
 for $T_p>1$ GeV the momentum $Q_1$ is large, $Q_1> 0.6$ GeV.
 This question is discussed in detail  in the Appendix  on the basis of
 the analytical expression for the np-transfer amplitude in the S-wave
 approximation.

 The contribution of the channel $\nu=1$ of the wave function to the
 $~^3He$ charge formfactor, $F_{ch}(\Delta)$, is shown  in Fig.
 \ref{figffhe3} by curve 2. Curve 3 in Fig. \ref{figffhe3} shows the result
 for $F_{ch}(\Delta)$ obtained with the function ${\widetilde{\chi}}_1(p_1)$
 instead of ${\chi}_1(p_1)$ and with ${\widetilde {\varphi}}_1(q_{23})$
  instead of  ${\varphi}_1(q_{23})$. One can
 see from this picture that the relative contribution of the channel
 $\nu=1$ is maximal at transferred momenta $\Delta > 1.5 $ GeV/c, but it is
 an order of magnitude smaller in comparison
 with the full result shown by curve 1 in Fig. \ref{figffhe3}.
 Moreover, the contribution obtained with the
 functions  ${\widetilde {\varphi}}_1(q_{23})$ and
 ${\widetilde{\chi}}_1(p_1)$, dominating in the cross section of
 the backward elastic  $p~^3He$-scattering, is negligible at all transferred
 momenta. Obviously it is connected to the fact that at high values
  $\Delta$ the charge formfactor $F_{ch}(\Delta)$ involves the high
 momentum components of the $^3He$ wave function associated both with the
 relative momentum  $q_{23}$  and  the momentum  $p_1$.

 Third, we have found numerically that the contribution
 of the OPE mechanism without taking into account rescatterings is in
 agreement with the experimental data at $T_p=0.5-1.3$ GeV (curve 4
 in Fig.\ref{fig5}). However, the ST cross section calculated in
 the Born approximation is by factor  $\sim 20-30$ larger than the
 OPE contribution  at $T_p>0.8$ GeV (curve 1 in Fig.\ref{fig5}).
\begin{figure}{}
\mbox{\epsfig{figure=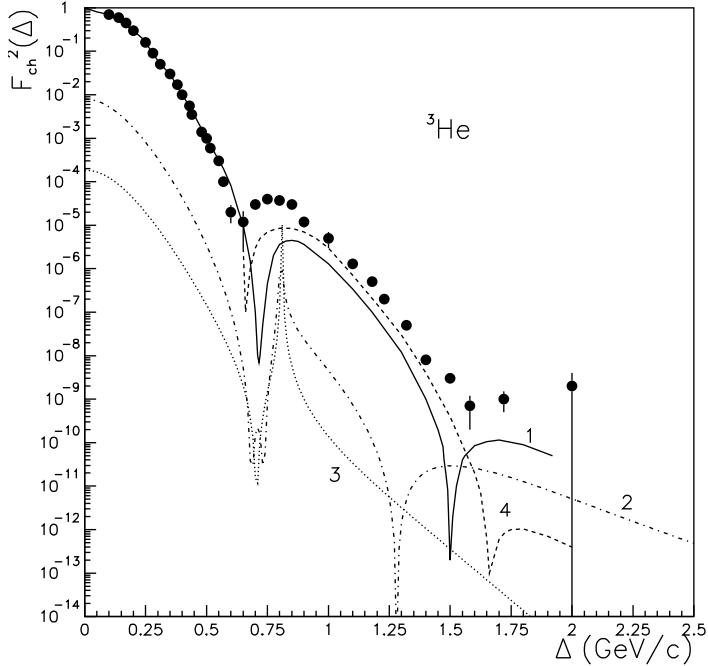,height=0.45\textheight, clip=}}
\caption{
 The charge formfactor of the  $^3He$ nucleus
 calculated  in the impulse approximation using dufferent assumptions
 about the $^3He$ wave function. Curve
 1 -- from \protect\cite{hgs} with  the three-body wave function of the
 $^3He$ nucleus;
 curve 2 -- with  only one channel $\nu=1$;
 curve 3 -- the same as  curve 2 but with
  ${\widetilde{\chi}}_1(p_1)$ instead of  ${\chi}_1(p_1)$ and
 ${\widetilde {\varphi}}_1(q_{23})$  instead of
 ${\varphi}_1(q_{23})$;
 curve 4 -- with the  $d+p$ configuration  in
Eqs.(\protect\ref{overlap}),
(\protect\ref{normawfhe3}).
   The circles ($\bullet $ ) are   experimental data from
 \protect\cite{arnold}.
}
\label{figffhe3}
\end{figure}

 The numerical results show that in the interval of initial
 energies 0.5- 2.5 GeV the contribution of the D-wave of the $d+p$ channel
 of the $~^3He$ wave function to the OPE
 amplitude is very small in comparison with the S-wave contribution.
 In fact the following
 numerical relations take  place for the formfactors defined by Eq.
 (\ref{formfact}):
$$|{\cal F}_2({\widetilde p})|\sim 0.1 |{\cal F}_0({\widetilde p})|,\,\,
|{\cal W}_{12}({\widetilde p},{\widetilde \delta})|\sim 0.1
|{\cal W}_{10}({\widetilde p}, {\widetilde \delta })|,\, \,
|{\cal W}_{32}({\widetilde p},{\widetilde \delta})|\sim 0.5 |{\cal W}_{12}
({\widetilde p}, {\widetilde \delta })|,$$
$$|{\cal F}_2({\widetilde p})|\sim
|{\cal W}_{32}({\widetilde p},{\widetilde \delta})|,\,
|{\cal F}_0({\widetilde p})|\sim 0.3
 |{\cal W}_{10}({\widetilde p}, {\widetilde \delta })|.$$
 The differential cross section of the process
 $pd\to ~^3He\pi^0$ at the $\pi$-meson scattering angle
 $\theta _{c.m.}=180^o$ is taken from the experimental data
  \cite{berth85}. The wave functions $U_0(r)$ and $U_2(r)$
 describing the relative  motion in the channel  $~^3He\to d+p$
 according to Eq. (\ref{overlap}) are parametrized here in the following form
\begin{equation}
\label{wfparam}
U_0(r)=\sum_{i=1}^5 S_i\exp{(-\kappa_i\,r^2)},\,\ \ \
U_2(r)=\sum_{i=1}^5 D_i r^2\exp{(-\lambda_i r^2)}.
\end{equation}
The numerical coefficients  $S_i,\ \kappa_i,\ D_i,\ \lambda_i$
are given in Table 1.
 The  $d+p$-configuration of the $~^3He$ nucleus
 described by   Eqs. (\ref{overlap}, \ref{normawfhe3})
 seems  reasonable enough for the evaluation of the OPE amplitude
 because this
 configuration  approximates  the $^3He$ charge formfactor
  properly in the wide region of  transferred momenta
 $\Delta=0\div 1.5 GeV/c$ (Fig.\protect\ref{figffhe3}).

\begin{figure}{}
\mbox{\epsfig{figure=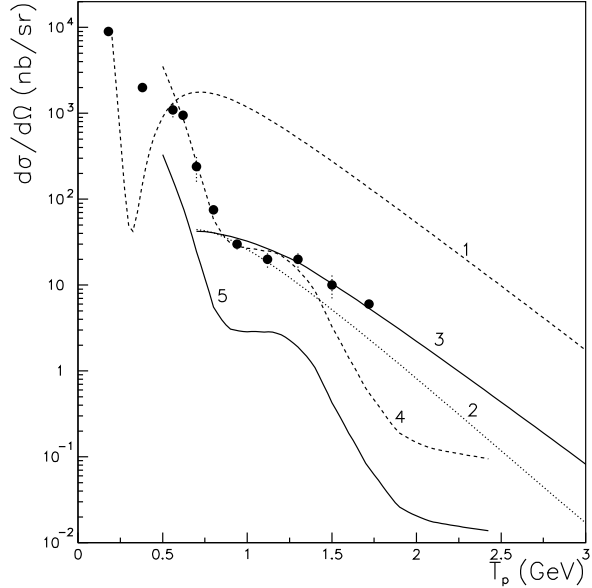,height=0.40\textheight, clip=}}
\caption{
 The same as in Fig.4
 but curves show the results
 of calculcalition with OPE and np-pair transfer mechanisms:
 1 --  Born approximation for the amplitude in Eq. (1)
 with the $~^3He$ wave function from [28]
,
 2 -- the same as curve 1 but with the deuteron w.f. $u(q_{23})$
 instead of  ${\varphi}_1(q_{23})$ and ${\varphi}_2(q_{23})$;
 3  -- the same as curve 1 but with allowance for rescatterings in
 the initial  and final states;
 4 -- OPE in the Born approximation,
 5 -- OPE with  rescatterings.
}
\label{fig5}
\end{figure}
 Taking into account rescatterings in the initial
 and final states we find that the
 cross section calculated with OPE mechanism decreases by one order
 of magnitude and becomes considerably  lower than the
 experimental data (see Fig. \ref{fig5}). The cross section
 of the $p~^3He\to ~^3Hep$ process for $T_p< 1 GeV$ is likely to be
 defined mainly by
 the multistep $pN$-scattering mechanisms discussed in Refs.\cite{pl,ls}
 including the heavy stripping mechanism \cite{gurvitz} -\cite{sherif} also.
 We stress that the high momentum components of the functions $\varphi_\nu$
 in Eq.(\ref{u3}) play the most important role in the competition between the
 OPE and ST mechanisms. One can see from Fig.\ref{fig3}, ({\it a}), that
 the high momentum component of the functions  $\varphi_\nu(q)$ is  richer
 in comparison  with the deuteron wave  function  $u(q)$, especially
 for $q> 0.5$ GeV/c.  This is a direct consequence of the
 fact that the $~^3He$ nucleus is more compact as compared with the deuteron.
 To compare with the $pd$-scattering, we performed the calculation of the
 ST cross section with the  S-component of the deuteron wave function
 $u(q)$  in Eq.(\ref{u3})  instead of the function $\varphi_\nu(q_{23})$
 for  $\nu=1,2$. As it seen from curve 2 in Fig. \ref{fig5}, in this case
 the ST cross section is by a factor $\sim $40 smaller than with the
 function $\varphi_\nu(q_{23})$ and close to the
 OPE cross section in the Born approximation.
 Note in this connection that in the
 $pd\to dp$ process the contribution of the neutron exchange mechanism
 in the Born
 approximation is not dominating \cite{azh} for $T_p> 1$GeV and  comparable
 with the OPE mechanism \cite{naksat, uz979}.
\begin{figure}{}
\mbox{\epsfig{figure=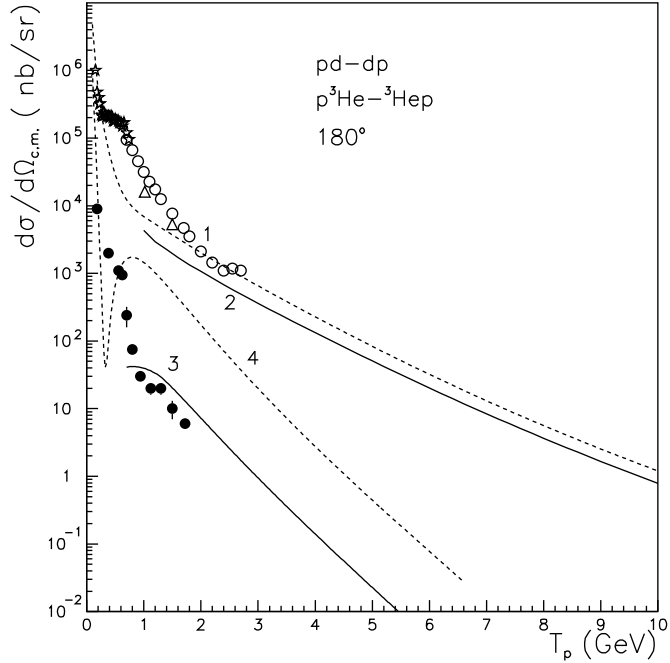,height=0.45\textheight, clip=}}
\caption{
 The differential cross sections of backward elastic
 $pd$- and  $p~^3He$-scattering at  $\theta_{c.m.}=180^o$ versus the
 kinetic energy of the initial proton  $T_p$.
 The curves show the results of calculations
 with the relativistic relative momenta in the vertices $d\to np$ and
 $^3He\to \{23\}+p$: curves  1 and  2 are taken  from
 Ref. \protect\cite{uz979} for
 the $pd\to dp$ process in the framework of the nucleon exchange
 mechanism  with RSC wave function of deuteron;
   3 and 4  refer to  the process $p~^3He\to ~^3He p $ for
 $np$-pair transfer mechanism. Dashed lines  (1 and 4) are the
 Born approximation,
 full lines (2 and  3)  take  rescatterings into account.
 The experimental points are taken from
   \protect\cite{berth} ($\bullet $),  \protect\cite{berthet82}
 ($\circ $),  \protect\cite{boudard} ($\star $),   \protect\cite{dubal}
 ($\triangle $).
}
\label{fig6}
\end{figure}
 This is one of reasons for
 a very nontrivial problem which arises when one attempts
 to extract a definite information about the high momentum components
 of the deuteron wave function from the experimental data on the
 $pd\to dp$ process.

 In Fig.\ref{fig6} the  cross sections of the  elastic $pd$-
 and $p~^3He$-scattering are compared with each other
 at the angle $\theta_{c.m.}=180^o$ as a functions
 of the initial energy $T_p$. One can see
 that with increasing initial energy the calculated cross section of the
  $p~^3He\to ~^3Hep$ process decreases more rapidly than the
 $pd\to dp$ cross section. The reason for this
 is the
 difference  between the deuteron and $~^3He$ masses. Owing to this fact
 the modulus of momentum  $Q_1=Q_0$ in Eq. (\ref{q1q0}) for the np-pair
 transfer mechanism of the  $p~^3He\to ~^3Hep$ process  increases
 with growing  $T_p$ essentially faster ( both in the nonrelativistic
 and relativistic kinematics) than  the relative momentum
 $q_{pn}$  in the vertex $d\to p+n$ of the pole diagram of the neutron
 exchange for the  $pd\to dp$ process. The results of calculations of these
 cross sections in the Born approximation
\begin{figure}{}
\mbox{\epsfig{figure=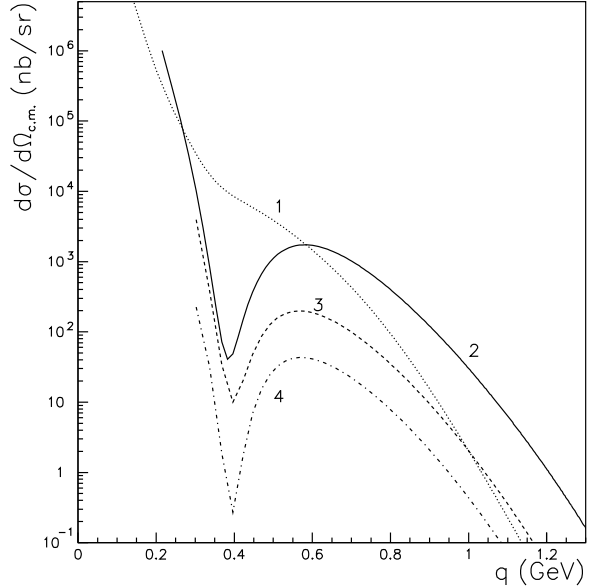,height=0.4\textheight, clip=}}
\caption{
 The same as in Fig.\protect\ref{fig6}, but versus the relative
 momentum  in the vertices $d\to n+p$ and $~^3He\to \{23\}+p$.
The curves show the
 results of calculations in the Born approxmation for the neutron
exchange in
 the $pd\to dp$ process ( curve 1), and for the np-transfer mechanism of
 the process $p~^3He\to ~^3He p$ (curves 2 - 4): 2 -- with
the $~^3He$ wave function from Ref. \protect\cite{hgs},
 3 -- with
the deuteron wave function $u(q)$ instead of
  $\varphi_1(q)$, 4 -- with the deuteron wave function $u(q)$ instead
  of $\varphi_1(q)$ and $\varphi_2(q)$.
}
\label{fig7}
\end{figure}
 are presented in
 Fig. \ref{fig7} versus the relative momentum
 $q_{pn}$  (for the   $pd\to dp$ process) and the momentum
 $Q_1=Q_0$ (for the  $p~^3He\to ~^3Hep$ process ).

 One can see from this figure that, in contrast to the $T_p$-dependence,
 with increasing $q$
 the $pd\to dp$ cross section as a function
 of the internal  relative  momentum $q$ decreases  more rapidly than
 the  $p~^3He\to ~^3Hep$ cross section.
 These cross sections  are
 equal to each other at the relative momentum
 $\sim 0.6 $GeV/c, which corresponds to the kinetic energy
 $T_p=2.65 $GeV in the   $pd\to dp$ process and  $T_p=0.8 $GeV in the
  $p~^3He\to ~^3Hep$ process. With  increasing the momenta up to
  $q\sim 1 $GeV/c the cross section of the $pd\to dp$ process
  becomes  an order of magnitude smaller than  the  $p~^3He\to ~^3Hep$
 cross section. At this point  the kinetic energy of incident
 proton equals to  9.3 GeV in  $pd-$ and  2.8 GeV in  $p~^3He-$ collision.
 As it seen from Fig.\ref{fig7}, after substitution of the deuteron
 S-wave function  $u(q)$ into Eq.(\ref{u3}) instead of the functions
 $\varphi_1$ and $\varphi_2$ the cross section of the $pd\to dp$ process
    decreases still faster than the cross section
 of the   $p~^3He\to ~^3Hep$ process.

  The above  performed comparison shows that an experimental investigation
 of the backward elastic $p~^3He$ scattering at energies of incident
 protons  $T_p\sim 2.5 $GeV can  give the unique information about
 the off-energy shell NN-interaction which might  be reached in
 the $pd-$ collision only at more high initial energies $\sim 9 $ GeV.

 The role of relativistic effects is estimated here by means of
 replacement of  the nonrelativistic momenta
 $Q^{nr}_1=Q^{nr}_0$ defined in Eqs.(\ref{q1q0}) with
 the corresponding relativistic ones $Q^{rel}_1=Q^{rel}_0$
 (where $Q^{rel} < Q^{nr}$) defined by Eq. (79) in Ref.\cite{[2]}.
 The results of calculations are shown in Fig. \ref{fig8}. As it seen
 from this figure, a such replacement turns out to be insignificant
 up to the initial  energy
  $T_p\sim 1 $ GeV, in spite of enough large magnitude of the nonrelativistic
 momentum $Q_0=Q_1\sim 0.6$ GeV/c at this  energy  for the scattering angle
 $\theta _{c.m.}=180^o$. With increasing the energy above  1  GeV
 the relativistic result for the cross section becomes considerably higher
  than the nonrelativistic one, thus at $T_p=3$ GeV the
 corresponding difference
\begin{figure}{}
\mbox{\epsfig{figure=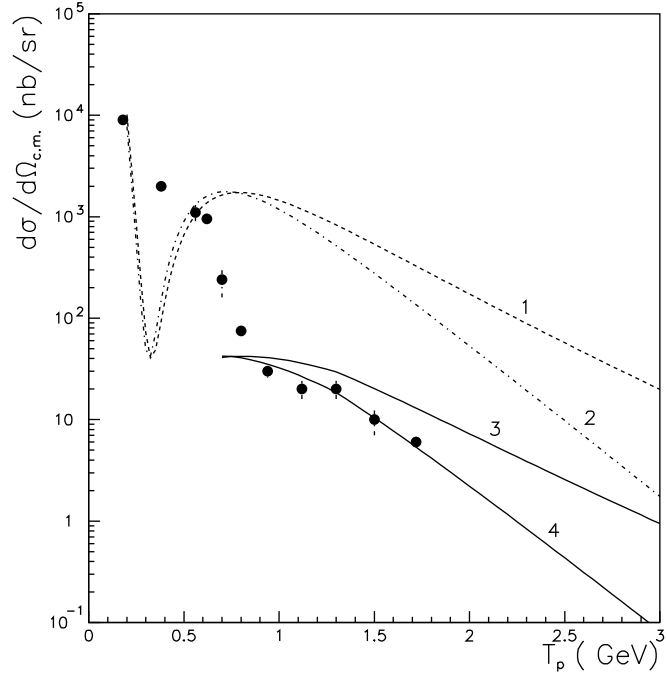,height=0.45\textheight, clip=}}
\caption{
 The same as in Figs.\protect\ref{fig4} -
 \protect\ref{fig6}, but calculated
 with the relativistic
 momenta   $Q_0, Q_1$ (curves  1 and 3) according to Ref.
 \protect\cite{[2]}
 and nonrelativisic momenta from Eq. (\protect\ref{q1q0})
(curves 2 and 4): 1,2 -- Born approximation; 3,4 -- with rescatterings
taken into account.}
\label{fig8}
\end{figure}

 is about an order of  magnitude. At this energy the nonrelativistic
 momentum $Q_1=Q_0$ takes the value $\sim $1.05 GeV/c. The relation
 between the relativistic and nonrelativistic result is not changed
 by the rescatterings.
 Therefore, in complete future analysis of this process one  should
 take into account relativistic effects in a consistent way.
 Note, that in the present work the agreement with experiment is better
 for the nonrelativistic calculations than for the relativistic ones.
  Perhaps, it is connected  to the fact that the RSC wave function
 \cite{bkt} is used here for the $~^3He$ nucleus.  The RSC potential
 provides for the wave functions of the lightest nuclei too intensive
 high  momentum components in comparison with  other realistic potentials
 like the  Paris potential (see, for example, \cite{iuzs89}).

\begin{figure}{}
\mbox{\epsfig{figure=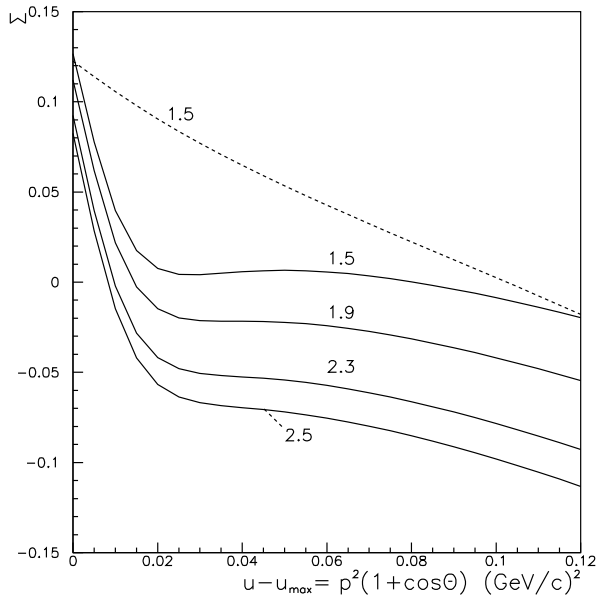,height=0.4\textheight, clip=}}
\caption{
 The spin-spin correlation parameter for the process
${\vec p} ~^3{\vec He} \to ~^3Hep$ as a fubction of $|u-u_{max}|$.
Full curves show the results of calculations taking into account
 rescatterings  in the initial and final states for different
 energies of the incident proton (GeV) shown near the curves;
 the dashed curve is the Born approximation for 1.5 GeV
}
\label{fig9}
\end{figure}

 The numerical results  for the parameter $\Sigma $ obtained
 with allowance for two channels $\nu=1$ and $\nu=2$ in the  $~^3He$
 wave function  are presented  in Fig. \ref{fig9} versus  the variable
 $|u-u_{max}|=2p^2(1+\cos{\theta_{c.m.}})$, where
 $p$ is the proton momentum in the $p+~^3He$ c.m.s. and $\theta_{c.m.}$ is
 the scattering angle. One can see  from this figure that at
 $ \theta _{c.m.}=180^o$   and $T_p\sim 1-2.5$ GeV the value $\Sigma$ is
 about $\sim 0.1 -0.15$ independently of the initial energy.
 Rescatterings in
 the initial and final states modify the form of angular dependence
 but do not change the energy dependence  at $\theta_{c.m.}=180^o$.
 The similar behaviour displays the spin averaged cross section
 \cite{[2]}.

\section{Conclusion}

 The question about a presence of nonnucleon degrees of freedom in the
 structure of the lightest nuclei at short NN-distances can be reformulated
 in other words in the following way. Up to what maximal values of relative
 momenta between nucleons inside a nucleus does the latter demonstrate the
 properties of the system which consists of nucleons with frozen
 internal degrees of freedom interacting by means of  realistic
 NN-potentials defined from the NN-phase shifts data?

  In this work the remarkable sensitivity of the cross section of backward
 elastic $p~^3He$-scattering to the high momentum components of the
 $~^3He$ wave function in the S-wave channel is found
 for energies above 1 GeV.
  The total dominance of nucleon degrees  of freedom
 in the $~^3He$ nucleus is demonstrated at these kinematical conditions.
 It is shown that  the backward elastic $p~^3He$-scattering advantageously
 differs  in this respect from the backward  elastic  $pd$-scattering.
 Some arguments are given to show that  this feature of the
 $p~^3He\to ~^3Hep$ process is connected with the
  high momentum component of the $~^3He$ wave function, which is
 more intensive in comparison with the deuteron wave function.
 Since the mechanism of the np-pair transfer describes  the available
 experimental data in the interval of incident energies 0.9-1.7 GeV
 satisfactorily, there is a reason to measure the cross section
 at higher energies in order to enlighten the validity  of phenomenological
 NN-potentials  in describing the structure of lightest nuclei at high
 relative  momenta of nucleons.

 {\bf Acknowledgements}.

 I am thankful to Prof. V.I. Komarov for discussions.
 This work was supported in part by the Russian Foundation for Basic Research
 (grant $N^o$ 96-02-17215).

\section*{Appendix }
 Here are presented the formulas for the amplitude of the np-pair
 transfer taking into account the  first two channels $\nu=1,2$
in the $~^3He$ wave function. The function given by Eq.(\ref{u3})
was approximated in  \cite{hgs}  by the sum of Yukawa terms.
Using the parametrization \cite{hgs} we found the
 gaussian parametrization
for the functions $f_\nu (p)=\{\phi_\nu(p),\chi_\nu(p)\}$,
\begin{equation}
\label{p1}
\varphi(q)=\sum_i G_i\exp{(-\alpha_iq^2)},
\ \ \ \chi(p)=\sum_j F_i\exp{(-\beta_ip^2)}
\end{equation}
 with the coefficients given in Table 2. These coefficients
 were found by means the minimization of the difference between the integrals
\begin{equation}
\label{svertka}
I({\bf Q})= \int f_\nu({\bf q})f_\nu ({\bf q}+{\bf Q})d{\bf q},
\end{equation}
 calculated in the interval  $Q=0\div 13 fm^{-1}$ with the functions
 $f_\nu (p)=\{\phi_\nu(p),\chi_\nu(p)\}$ from Ref. \cite{hgs} on the
 one side and with the gaussian  parametrization in
 Eq. (\ref{p1})  on the other side.
\eject
{\bf Tabl.1 Coefficients for the expansions in Eq. (\ref{wfparam}) }

\begin{tabular}{cc|cc}
\hline
\multicolumn{4}{r}{}\\
\multicolumn{1}{c}{ $S_i,\, fm^{-3/2}$ } &
\multicolumn{1}{c|}{ $\kappa _i,\, fm^{-2}$ } &
\multicolumn{1}{c}{ $D_i,\, fm^{-7/2}$ } &
\multicolumn{1}{c}{ $\lambda _i,\, fm^{-2}$ }\\
\multicolumn{4}{r}{}\\
\hline
\multicolumn{4}{r}{}\\
1.80112E-02 & 2.15766 E-02 &  -1.93862E-03 &  9.83826 E-02 \\
2.13255E-01 & 8.35379 E-02 &  -1.58838 E-02 & 3.18527 E-01 \\
9.00237E-02 & 1.27578 E-01 &  -3.11061 E-02 & 6.43963 E-01 \\
3.23190E-01 & 3.26778 E-01 &  -3.83184 E-02 & 1.19183 E+00 \\
-2.17017E-01 & 1.06206 E+00 & -9.57312 E-02 & 4.47721 E+00 \\
\multicolumn{4}{r}{}\\
\hline
\label{tabl1}
\end{tabular}

{\bf Tabl.2 Coefficients for the expansions in Eq. (\protect\ref{p1})}

\begin{tabular}{lllll}
\hline
\multicolumn{5}{r}{ }\\
\multicolumn{1} {c}{$\nu $ } & \multicolumn{2} {c}{ $\phi_\nu$ \ \ \ \ \ \ } &
\multicolumn{2} {c}{\ \ \ \ \ \ \  $\chi_\nu $ \ \ \ \ \ }\\
 \cline {1-3}  \cline {4-5}
\multicolumn{5}{r}{ }\\
\multicolumn{1}{c}{}&\multicolumn{1}{l} {$G_i,\, fm^{\frac{3}{2}}$} &
\multicolumn{1}{c} {$\alpha _i,\, fm^{2}$} &
\multicolumn{1}{c} {$F_i,\, fm^{\frac{3}{2}}$} &
\multicolumn{1}{r}{ $\beta _i,\, fm^{2}$}\\
\hline
\multicolumn{5}{r}{ }\\
  &  2.413615       & 5.535106        & 4.20690      & 7.62335      \\
  & -1.299993  E-01 & 1.060713 E-01   & 2.59354      & 2.37678       \\
1 &  4.118231  E-02 & 4.555083 E-03   & 4.8189 E-01  & 9.18116 E-01  \\
  &  1.762614  E-00 & 1.611916 E+00   &-2.45993E-02  & 1.60613 E-01  \\
  &  7.491376  E-01 & 5.466668 E-01   &5.21600  E-03 & 1.40849 E-02  \\
  & -4.00000   E-02 & 4.526665 E-03   &-5.67292 E-03 & 1.03918 E-02  \\
  & -3.451822  E-02 & 5.023824 E-02   & 2.17920 E-03 & 7.19787 E-03  \\
\multicolumn{5}{r}{ }\\
  &-2.54039     & 6.55162      &  6.51121       & 9.38385     \\
  &-1.93783     & 1.74914      &  2.59354       & 3.04139     \\
2 &-7.74249 E-01 & 5.72939 E-01 &  5.45100 E-01 & 1.22803      \\
  & 1.12706 E-01 & 1.15040 E-01 & -2.40357 E-02 & 1.85314 E-01 \\
  & 3.09857 E-02 & 5.11790 E-02 &  1.56646 E-03 & 1.84247 E-02 \\
  & -6.4000 E-03 & 3.94752 E-02 & -2.61864 E-03 & 1.03918 E-02 \\
  & 1.18765 E-03 & 6.58383 E-03 & 2.17920 E-03 & 9.12018 E-03 \\
\multicolumn{5}{r}{ }\\
\hline
\label{tabl2}
\end{tabular}


 Using the gaussian parametrization for the functions
 $\varphi(q)$ and $\chi(p)$ we find the following expression for the
 integrals in Eq.(\ref{int})
$$I({\bf Q}_1, {\bf Q}_0)=\frac{1}{(4\pi )^2}\int d^3q (q^2+M^2)
\varphi_{\nu'}(\gamma'{\bf q}+\delta '{\bf Q}_0)
\varphi_{\nu}(\gamma { \bf q}+\delta {\bf Q}_1)$$
$$\chi_\nu(a{\bf q}+b {\bf Q_1})\chi_{\nu'}(a'{\bf q}+b' {\bf Q_0})
= \frac{1}{(4\pi )^2}\sum _{i,j,k,l}G_iG_jF_kF_l
\left (\frac{\pi}{D}\right )^{3/2}$$
\begin{eqnarray}
\label{p2}
\times \left \{ M^2+D^{-1}\left (\frac{3}{2}+B^2\, D^{-1}\right )\right \}
\exp{\left [\frac{B^2}{D}-C\right ]},
\end{eqnarray}
where
\begin{eqnarray}
\label{p3}
C=\left [\alpha _i (\delta ')^2+\alpha _j\delta ^2+\beta_k b^2+\beta _l(b')^2
\right ] Q^2,\nonumber \\
D=\alpha_i (\gamma')^2+\alpha_j\gamma^2 +\beta_k a^2+\beta_l(a')^2,\nonumber \\
B^2=\left (\alpha_i\gamma'\beta '+\beta_l a'b'\right )^2Q_0^2+
\left (\alpha_j\gamma \delta+ \beta _k ab\right )^2 Q_1^2+\nonumber \\
+2{\bf Q}_1{\bf Q}_0 \left (\alpha_i\gamma'\beta '+\beta_l a'b'\right )
\left (\alpha_j\gamma \delta+ \beta _k ab\right ),
\end{eqnarray}
 In Eq.(\ref{p2}-\ref{p3}) the summation over the indices  $i,\,j,\, k, \, l $
 refers to the expansions in Eq.(\ref{p1}) for the functions
  $\varphi_{\nu '},\,\varphi_{\nu },\, \chi_{\nu },\,
\chi_{\nu '}$, respectively.

From Eq. (\ref{momenta}) one finds for the ST mechanism :
\begin{eqnarray}
\label{p4}
  \gamma '=-\frac{1}{2},\, \delta= -\frac{3}{4}; \gamma=-\frac{1}{2},
\delta=\frac{3}{4},\nonumber \\
a'=1,\, b'=-\frac{1}{2}; \,  a=-1,\, b=-\frac{1}{2}.
\end{eqnarray}
Therefore, at the scattering angle  $\theta_{c.m.}=180^0$
 ({\it et id.} ${\bf Q}_1=-\bf {Q}_0)$ the expression in the exponent
 for the ST amplitude in Eq.(\ref{p2}) takes the form
\begin{equation}
\label{p5}
\left [\frac{B^2}{D}-C\right ]_{ST}=
-{\bf Q}_0^2\left \{ \frac {(\alpha_i+\alpha_j)(\beta_k+\beta_l)}
{ \frac{1}{4}(\alpha_i+\alpha_j)+(\beta_k+\beta_l)} \right \}.
\end{equation}

 One can find the  following three conditions for which the absolute
 magnitude  of the value in the right hand side of Eq.(\ref{p5}) has
 a minimum at  ${ Q}_0= { Q}_1=const$.

A) In the case  $\alpha_i+\alpha_j\ll \beta_k+\beta_l\sim 1$
 one finds from Eq.(\ref{p5}) the relation \\
$-\frac{1}{{\bf Q}^2}_0\left [\frac{B^2}{D}-C\right ]_{ST}
\sim (\alpha_i+\alpha_j)(1-\frac{1}{4}\frac{\alpha_i+\alpha_j}
{\beta_k+\beta_l})\ll 1$. Due to the relation
$\beta_k+\beta_l\sim 1$ the product of   pre-exponentials
  $F_kF_l$ in Eq.(\ref{p2}) is large  according to Table 2 which
 gives the correspondence between $\beta_i$ and $F_i$.

B) $ \beta_k+\beta_l\ll \alpha_i+\alpha_j\sim 1$. It corresponds to
 the relation
 $-\frac{1}{{\bf Q}^2}_0\left [\frac{B^2}{D}-C\right ]_{ST}
\sim 4 (\beta_k+\beta_l)$.

C) $ \beta_k+\beta_l\sim \alpha_i+\alpha_j\ll 1$.
 In this case one finds  $-\frac{1}{{\bf Q}^2}_0\left [\frac{B^2}{D}-C\right ]_{ST}
\sim 4 (\beta_k+\beta_l)$, however all pre-exponentials
  $G_i,\,G_j,\, F_k,\, F_l$ in Eq.(\ref{p2}) are small (see Table 2).

 It is obvious, that only in the case A) the exponent takes the minimal
 value which corresponds to the maximal value of the ST amplitude.
In this case the first condition  $\alpha_i+\alpha_j\ll  1$ means that the
 high momentum components of two functions $\varphi_{\nu'}(q)$ and
 $ \varphi_{\nu}(q)$ are involved in the ST-amplitude, whereas the
 second condition $ \beta_k+\beta_l\sim 1$ corresponds to soft momenta
($p\sim 1/\sqrt{\beta_k},\, 1/\sqrt{\beta_l}$) in the other two
 functions $\chi_{\nu'}(p),\, \chi_{\nu}(p)$
\footnote{Note that using the harmonic oscillator
translationally-invariant shell model wave function for the  $~^3He$ nucleus
\cite{neudsm69}  one keeps  only one gaussian term in Eqs. (\ref{p1})
 with the exponents $\alpha $ and $\beta $ related as
  $\beta=\frac{3}{4}\alpha$. In this case
 the positive term  $B^2/D$ in the  exponent of Eq.(\ref{p2}), which moderates
 the decrease of the amplitude with increasing  $Q_0$,
 vanishes
 for the ST mechanism ( $B^2/D=0$). As a result, the ST mechanism
 reduces to the deuteron exchange mechanism.}.

 Similarly, one can find for the NST mechanism :
\begin{eqnarray}
\label{p6}
  \gamma '=-\frac{1}{2},\, \delta= -\frac{3}{4}; \gamma=-\frac{1}{2},
\delta=-\frac{3}{4},\nonumber \\
a'=1,\, b'=-\frac{1}{2}; \,  a=1,\, b=-\frac{1}{2}.
\end{eqnarray}
\begin{equation}
\label{p7}
-\frac{1}{{\bf Q}_0^2}\left [\frac{B^2}{D}-C\right ]_{NST}=
\left \{ \frac {(\alpha_i+\alpha_j)(\beta_k+\beta_l)
+\left [\frac{3}{2}\alpha_i\alpha_j +2 \beta_k \beta_l\right ]}
{ \frac{1}{4}(\alpha_i+\alpha_j)+(\beta_k+\beta_l)} \right \}.
\end{equation}
 Due to an additional term in the square brackets in the
 numerator on the right hand side of Eq. (\ref{p7}) the exponent
 for the NST mechanism is always larger in absolute value
 than the exponent for the ST mechanism. For this reason
 the contribution of the  NST mechanism to the cross section is
 considerably smaller in comparison with  the ST mechanism.

The IPT amplitude is defined by the integral in Eq. (\ref{p2}) under
 the following conditions \cite{[2]}
\begin{eqnarray}
\label{p8}
  \gamma '=1,\, \delta '=0; \gamma=-\frac{1}{2},
\delta=\frac{3}{4},\nonumber \\
a'=0,\, b'=1; \,  a=-1,\, b=-\frac{1}{2};
\end{eqnarray}
\begin{equation}
\label{p9}
-\frac{1}{{\bf Q}_0^2}\left [\frac{B^2}{D}-C\right ]_{IPT}=
\left \{\beta_l +\frac{\alpha_j\left (\beta_k+\frac{9}{4}\alpha_i
\right ) +\frac{1}{4}\alpha_i^2}{\alpha_i+\frac{1}{4}\alpha_j+\beta_k }
\right \}.
\end{equation}
It is easy to find that the minimum of the right-hand side in Eq.(\ref{p9})
 takes place for
$\beta_l\ll 1,\, \alpha_i\ll \beta_k, \, \alpha_j\ll \beta_k$.
This means that the main contribution  to the
IPT amplitude comes from the high momentum components of three functions
$\chi_{\nu'},\, \varphi_{\nu'}$ and $ \varphi_{\nu}$
simultaneously.

\eject

\eject
\end{document}